\documentclass[11pt]{JHEP3}

\usepackage{epsfig}
\usepackage[latin1]{inputenc}
\usepackage{bbm,amsfonts}
\usepackage{graphicx}
\usepackage{amssymb,amsmath}
\usepackage{epsfig,multicol,bbm}
\usepackage{slashed}

\newcommand\fverb{\setbox\fverbbox=\hbox\bgroup\verb}
\newcommand\fverbdo{\egroup\medskip\noindent%
            \fbox{\unhbox\fverbbox}\ }
\newcommand\fverbit{\egroup\item[\fbox{\unhbox\fverbbox}]}
\newbox\fverbbox

\title{Notes on Emergent Gravity}

\author{Sunggeun Lee, Raju Roychowdhury and Hyun Seok Yang \\
 Center for Quantum Spacetime, Sogang University, Seoul 121-741, Korea\\
E-mail:
\email{sglkorea@hotmail.com, raju.roychowdhury@gmail.com, hsyang@sogang.ac.kr}}

\received{\today}       
\accepted{\today}       

\preprint{arXiv:1206.0678}

\abstract{Emergent gravity is aimed at constructing a Riemannian geometry from
U(1) gauge fields on a noncommutative spacetime. But this
construction can be inverted to find corresponding U(1) gauge fields
on a (generalized) Poisson manifold given a Riemannian metric $(M,
g)$. We examine this bottom-up approach with the LeBrun metric which
is the most general scalar-flat K\"ahler metric with a U(1) isometry
and contains the Gibbons-Hawking metric, the real heaven as well as
the multi-blown up Burns metric which is a scalar-flat K\"ahler
metric on $\mathbb{C}^2$ with $n$ points blown up. The bottom-up
approach clarifies some important issues in emergent gravity.}


\begin{document}


\section{Introduction}

Recently the correspondence between noncommutative (NC) U(1) gauge
theory and gravity has evolved at large in the context of emergent
gravity. See
\cite{review1,review2,lee-yang,review4,hsy-jpcs12,review6,review7} for recent reviews.
The idea of emergent gravity is the following: Suppose that U(1)
gauge theory is defined on a symplectic manifold $(M, B)$ where $B$
is a nondegenerate, closed two-form on a smooth manifold $M$. Indeed
one can consider the symplectic two-form $B$ as a field strength of
vacuum gauge fields which take the form $A_\mu^{(0)} = -
\frac{1}{2}B_{\mu\nu} y^\nu$ on a local Darboux chart. Let us
introduce dynamical gauge fields fluctuating around the backbround
$B = d A^{(0)}$. The resulting field strength is given by $\mathcal
F= B + F$ where $F=dA$ is the curvature two-form of the dynamical
gauge field $A$. Note that $d\mathcal F=0$ due to the Bianchi
identity $dF=0$ and ${\mathcal F}$ is invertible unless $\det(1+
B^{-1}F)=0$. Therefore ${\mathcal F}= B + F$ is again a symplectic
structure on $M$ and so the dynamical gauge fields defined on a
symplectic vacuum $B$ manifest themselves as a deformation of the
symplectic structure \cite{hsy-ijmp09,hsy-jhep09}.

One may introduce local coordinates $X^a, \; a = 1, \cdots, 4$, on a
local chart $U \subset M$ where the symplectic structure ${\mathcal
F}$ is represented by
\begin{equation}\label{symp-omega}
    {\mathcal F} = \frac{1}{2}\Big( B_{ab} + F_{ab}(X) \Big) dX^a \wedge
    dX^b.
\end{equation}
But one can introduce another coordinates, say $y^\mu$, on the same
local patch $U \subset M$ which are diffeomorphic to $X^a$, i.e.
$X^a = X^a (y)$. Now one can ask an interesting question whether it
is possible to find a coordinate transformation $f: X
\mapsto y=y(X)$ in order to eliminate the electromagnetic force
$F=dA$ in the symplectic structure ${\mathcal F} = B + F$. In other
words, one may try to find a local coordinate transformation $f: X
\mapsto y=y(X)$ such that the symplectic structure ${\mathcal F}$ in
(\ref{symp-omega}) on $U
\subset M$ becomes
\begin{equation}\label{darboux-frame}
    {\mathcal F}|_U = \frac{1}{2} B_{\mu\nu} dy^\mu \wedge
    dy^\nu.
\end{equation}
Remarkably, the Darboux theorem or the Moser lemma in symplectic
geometry \cite{sg-book1,sg-book2} says that it is always possible to
find such a local coordinate transformation as long as the space $M$
admits a symplectic structure. If so, it is immediate to see from
(\ref{symp-omega}) that the so-called Darboux coordinates $y^\mu$
will obey the following relation \cite{cornalba,jur-sch}
\begin{equation}\label{darboux-tr}
 \Big(B_{ab} + F_{ab}(X) \Big)\frac{\partial X^a}{\partial y^\mu}
 \frac{\partial X^b}{\partial y^\nu} = B_{\mu\nu}.
\end{equation}

By taking the inverse of (\ref{darboux-tr}), one can rewrite it in
the form
\begin{equation}\label{poisson-tr}
    \Theta^{ab} (X) \equiv \Big( \frac{1}{B+F} \Big)^{ab}(X) =
  \theta^{\mu\nu}  \frac{\partial X^a}{\partial y^\mu}
 \frac{\partial X^b}{\partial y^\nu} \equiv \{ X^a, X^b \}_\theta
 (y)
\end{equation}
where $\theta \equiv \Big( \frac{1}{B} \Big) =
\frac{1}{2} \theta^{\mu\nu} \frac{\partial}{\partial y^\mu} \wedge
\frac{\partial}{\partial y^\nu} \in \Gamma(\wedge^2 TM)$ is a bivector field
that defines a Poisson structure on $M$. The Poisson structure
defines an $\mathbb{R}$-bilinear operation $\{-,-\}_\theta$, the
so-called Poisson bracket \cite{sg-book1,sg-book2}, given by
\begin{equation}\label{p-bracket}
    (f,g) \mapsto \{f, g\}_\theta = \theta(df,dg) = \theta^{\mu\nu}
    \frac{\partial f}{\partial y^\mu}
 \frac{\partial g}{\partial y^\nu}
\end{equation}
for smooth functions $f,g$. Let us represent the coordinate
transformation in the following form
\begin{equation}\label{coord-x}
    X^a(y) = y^a + \theta^{ab} \widehat{A}_b(y).
\end{equation}
Then (\ref{poisson-tr}) reads as \cite{liu,hsy-mpla06,ban-yan}
\begin{equation}\label{sw-map}
\Theta^{ab} (X) = \Big(\theta - \theta  \widehat{F} \theta
\Big)^{ab} (y) \quad \Leftrightarrow \quad
\widehat{F}_{\mu\nu} (y) = \Big(\frac{1}{1+ F \theta}
F\Big)_{\mu\nu} (X)
\end{equation}
where
\begin{equation} \label{nc-f}
 \widehat{F}_{\mu\nu}  = \partial_\mu \widehat{A}_\nu
- \partial_\nu \widehat{A}_\mu  + \{ \widehat{A}_\mu,
\widehat{A}_\nu \}_\theta.
\end{equation}
Once we know fluctuations described by $F=dA$, we can, in principle,
solve (\ref{sw-map}), known as the Seiberg-Witten map
\cite{ncft-sw}, to find the coordinate transformation (\ref{coord-x}) that
locally eliminates the electromagnetic force $F=dA$.

In the end we have arrived at an important result
\cite{hsy-ijmp09,hsy-jhep09} that the electromagnetic force can
always be eliminated by a local coordinate transformation as long as
U(1) gauge theory is defined on a symplectic manifold $M$ with
symplectic structure $B$. In other words, there exists an analogue
of the equivalence principle even for the electromagnetic force
whenever U(1) gauge fields have a vacuum condensate $\langle
\mathcal{A}_\mu (y) \rangle_{\mathrm{vac}} \equiv A_\mu^{(0)} (y) = -
\frac{1}{2}B_{\mu\nu} y^\nu$.\footnote{As will be shown later, the
equivalence principle for the electromagnetic force guarantees that
gravity can emerge from NC U(1) gauge theory \cite{hsy-jhep09}. It
turns out that the emergent gravity from NC U(1) gauge fields can be
formulated in a background independent way where no spacetime
structure is assumed but defined by the theory itself
\cite{review4,hsy-jpcs12}. Therefore one should not interpret the vacuum gauge
field $A_\mu^{(0)} (y) = - \frac{1}{2}B_{\mu\nu} y^\nu$ as an extra
background condensed on a pre-existing spacetime. The flat spacetime
(with Lorentz symmetry as an isometry) will emerge as a result of
the vacuum condensate and hence it does not break the Lorentz
symmetry \cite{hsy-jhep09}.} Consequently, U(1) gauge theory on a
symplectic manifold $(M,B)$ boils down to solving the Seiberg-Witten
map (\ref{sw-map}). If one has successfully solved (\ref{sw-map}) to
determine $\widehat{A}_\mu (y)$ (which will be identified with NC
U(1) gauge fields after quatization), all (at least local)
informations of electromagnetic fields on the symplectic vacuum $B$
are encoded into the coordinate transformation (\ref{coord-x}).

Since the coordinates $X^a(y)$ can be regarded as smooth functions
on $M$ and they are defined on a Poisson manifold $(M, \theta)$ as
was already implied by (\ref{poisson-tr}), one can define an adjoint
operation in the Poisson algebra:
\begin{equation}\label{gierbein}
    V_a (f) = \{ C_a, f \}_\theta
\end{equation}
where $f(y)$ is a smooth function and
\begin{equation}\label{nc-c}
C_a(y) \equiv B_{ab} X^b(y) = B_{ab} y^b + \widehat{A}_a(y)
\end{equation}
will be dubbed as ``symplectic gauge fields."\footnote{We have
observed in (\ref{darboux-tr}) that $X^a$s in (\ref{coord-x}) arise
as a local trivialization of line bundle $L \to U$ over a Darboux
chart $U$. Thus one can regard the symplectic gauge field $C_a$ as a
local section of the line bundle $L$ (or more precisely, a sheaf of
local functions).} The adjoint operation (\ref{gierbein}) satisfies
the Leibniz rule, i.e.,
\begin{equation}\label{leibniz}
\{ C_a, f \cdot g \}_\theta = \{ C_a, f \}_\theta \cdot g + f \cdot \{ C_a, g \}_\theta
\end{equation}
for any functions $f,g$ and thus $V_a$'s can be regarded as
derivations. In particular, $V_a$ can be identified with vector
fields on tangent bundle $TM \to M$, that is, $V_a \in \Gamma(TM)$.
Since the U(1) gauge fields $A_\mu (X)$ are encoded into the
coordinate transformations $\widehat{A}_\mu (y)$ via the Darboux
theorem and then mapped to vector fields in (\ref{gierbein}), the
U(1) gauge theory on symplectic manifold $(M, B)$ can now be
transformed into some geometry described by the vector fields $V_a$
\cite{lee-yang}.

In terms of local coordinates $y^\mu$ on a Darboux chart $U
\subset M$, the Hamiltonian vector fields $V_a\in \Gamma(TM)$
are given by
\begin{equation}\label{vector-v}
  V_a = V_a^\mu (y) \frac{\partial}{\partial y^\mu} \qquad
  \mathrm{with} \qquad V_a^\mu (y) = - \theta^{\mu\nu}
  \frac{\partial C_a(y)}{\partial y^\nu}.
\end{equation}
The emergent gravity is defined by identifying a map from the vector
fields in (\ref{vector-v}) to a gravitational metric given by
\begin{equation}\label{riem-metric}
    ds^2 = g_{\mu\nu}(x) dx^\mu \otimes dx^\nu = e^a \otimes e^a.
\end{equation}
This formulation of emergent gravity to define a gravitational
metric from symplectic gauge fields in (\ref{nc-c}) will be called
the top-down approach in comparison with the bottom-up approach to
identify symplectic gauge fields from a given gravitational metric.
In this paper we want to address the bottom-up approach of emergent
gravity. In this respect, we want to emphasize that the coordinates
$y^\mu$ are Darboux coordinates satisfying the relation
(\ref{darboux-tr}). But the metric (\ref{riem-metric}) has to
respect the general covariance and is represented in a general
coordinate system, denoted by $x^\mu$, which is not necessarily in
the Darboux frame.\footnote{For this reason, we will explicitly
distinguish Darboux coordinates $y^\mu$ in gauge theory and general
coordinates $x^\mu$ appearing in a gravitational metric.} One might
recall that there are many different coordinate systems to represent
the same metric. For example, the usual spherical coordinate
representation of Eguchi-Hanson metric \cite{eh-metric} is
equivalent to the two-center Gibbons-Hawking metric \cite{gh-metric}
by a coordinate transformation \cite{prasad} though their bare
appearance looks very different. Therefore, in order to identify a
gravitational metric from the vector fields $V_a$, it is convenient
to first perform a general coordinate transformation from $y^\mu$ to
$x^\mu$, i.e. $y^\mu \mapsto x^\mu=x^\mu(y) \in
\mathrm{Diff}(M)$ and represent the Poisson algebra
$\mathfrak{P}(M) = (C^\infty(M), \{-,-\}_\theta)$ in such a
coordinate system \cite{hsy-jhep09}. In order to clarify this point,
let us rewrite the Poisson bracket in (\ref{p-bracket}) in the
general coordinate system $\{ x^\mu \}$:
\begin{eqnarray}\label{px-bracket}
\{f, g\}_\theta &=& \theta^{\mu\nu}
    \frac{\partial f}{\partial y^\mu}
 \frac{\partial g}{\partial y^\nu} = \theta^{\mu\nu}
    \frac{\partial x^\rho}{\partial y^\mu}
 \frac{\partial x^\sigma}{\partial y^\nu} \frac{\partial f}{\partial x^\rho}
 \frac{\partial g}{\partial x^\sigma} \nonumber \\
 &=& \{x^\mu, x^\nu \}_\theta \frac{\partial f}{\partial x^\mu}
 \frac{\partial g}{\partial x^\nu} = \Theta^{\mu\nu} \frac{\partial f}{\partial x^\mu}
 \frac{\partial g}{\partial x^\nu} = \{f, g\}_\Theta
\end{eqnarray}
where
\begin{equation}\label{g-poisson}
\Theta^{\mu\nu} (x) \equiv \{x^\mu, x^\nu \}_\theta
\end{equation}
is the Poisson structure in the coordinate system $\{ x^\mu \}$.
Indeed the definition (\ref{g-poisson}) reduces to
(\ref{poisson-tr}) if one identifies $x^\mu = X^\mu$ and $x^\nu =
X^\nu$ and so the identity (\ref{px-bracket}) is a Poisson algebra
version of the Darboux theorem (\ref{darboux-tr}).

Let us define a vector field $X_f$ for a smooth function $f
\in C^\infty(M)$ in the general coordinate system $\{ x^\mu \}$ by
\begin{equation}\label{gh-vec}
    X_f(g) \equiv \{f, g\}_\Theta  \qquad \Leftrightarrow \qquad
    X_f^\mu (x) = - \Theta^{\mu\nu}(x) \frac{\partial f(x)}{\partial
    x^\nu}.
\end{equation}
Also we define a two-form
\begin{equation}\label{gsym-str}
\Omega = \frac{1}{2} \Omega_{\mu\nu}(x) dx^\mu \wedge dx^\nu
\end{equation}
uniquely determined by the Poisson tensor $\Theta^{\mu\nu} (x) =
(\Omega^{-1})^{\mu\nu}(x)$. The transformed Poisson bracket is then
represented by
\begin{equation}\label{t-poisson-bracket}
    \{f, g\}_\Theta = \Omega(X_f, X_g) = X_f(g) = - X_g(f)
\end{equation}
for $f, g \in C^\infty(M)$. The identity (\ref{px-bracket})
immediately shows that the Poisson algebra $\mathfrak{P}(M) =
(C^\infty(M), \{-,-\}_\Theta)$ has to obey the Jacobi identity. It
requires $\Omega$ to be a closed two-form, i.e. $d\Omega=0$ because
of the identity
\begin{eqnarray}\label{jacobi-gen}
  &&    \{ \{f, g\}_\Theta, h \}_\Theta + \{ \{g, h\}_\Theta, f \}_\Theta
  + \{ \{h, f\}_\Theta, g \}_\Theta \nonumber  \\
  &=& - \Big( X_h \big( \Omega(X_f, X_g) \big)
+ X_f \big( \Omega(X_g, X_h) \big) + X_g \big( \Omega(X_h, X_f)
\big) \Big) \nonumber  \\
  &=& - d\Omega (X_f, X_g, X_h) = 0.
\end{eqnarray}
In other words, $(M, \Omega)$ is also a symplectic
manifold.\footnote{\label{diff-darboux}This symplectic manifold can
be understood as follows. Consider an arbitrary split of the
electromagnetic field, $F = F_1 + F_2$, and suppose $\Omega = B +
F_1$ to be a primitive symplectic structure on $M$. One can consider
a coordinate transformation $\phi \in \mathrm{Diff}(M)$ such that
$\phi^*(\Omega+F_2) = \Omega$. Then $F_1=0$ where $\Omega=B$
recovers (\ref{darboux-tr}) whereas $F_2 = 0$ where $\phi =
\mathrm{identity}$ corresponds to commutative gauge theory.
It is also straightforward to check that $\{X^\mu, X^\nu \}_\Theta =
\Big(\frac{1}{\Omega+F_2} \Big)^{\mu\nu} =
\{X^\mu, X^\nu \}_\theta$.} Hence $\mathfrak{P}(M) = (C^\infty(M),
\{-,-\}_\Theta)$ is also a Lie algebra (called the Poisson-Lie
algebra of $(M, \Omega)$) and the mapping $\mathfrak{H}:
\mathfrak{P}(M) \to \mathfrak{X}(M)$ (where $\mathfrak{X}(M)$ is the
Lie algebra of vector fields of $M$) defined by $ f \mapsto X_f$ is
a Lie algebra homomorphism \cite{sg-book1,sg-book2}, i.e.,
\begin{equation}\label{g-lie-homo}
    X_{\{f, g\}_\Theta} = [X_f, X_g].
\end{equation}

According to (\ref{gh-vec}), the vector fields in a general
coordinate system for the symplectic gauge fields in (\ref{nc-c})
are defined by
\begin{equation}\label{ham-vc}
 V_a = V_a^\mu (x) \frac{\partial}{\partial x^\mu} \qquad
  \mathrm{with} \qquad V_a^\mu (x) = - \Theta^{\mu\nu} (x)
  \frac{\partial C_a(x)}{\partial x^\nu}
\end{equation}
and the symplectic gauge fields are assumed to take the form
\begin{equation}\label{symp-gen}
    C_a(x) = \Omega_{ab} (x) x^b + \widehat{\mathcal{A}}_a(x).
\end{equation}
Hence the Hamiltonian vector fields in (\ref{vector-v}) can be
transformed into the vector fields in (\ref{ham-vc}) by a general
coordinate transformation \cite{hsy-jhep09}:
\begin{equation}\label{diff-vec}
V_a^\mu (x) =
\frac{\partial x^\mu}{\partial y^\nu} V_a^\nu(y).
\end{equation}
Note that the components of the vector field $V_a \in \Gamma(TM)$
can be written in an inspiring form
\begin{eqnarray}\label{vec-comp}
 V_a^\mu (x) &=& V_a (x^\mu) = \{ C_a, x^\mu \}_\theta (y) \nonumber \\
 &=& \frac{\partial x^\mu(y)}{\partial y^a}
 + \{ \widehat{A}_a, x^\mu \}_\theta (y) \nonumber \\
 &\equiv & D_a x^{\mu},
\end{eqnarray}
where both $C_a(y)$ and $x^\mu(y)$ are regarded as functions of the
Darboux coordinates $y^\mu$. It has to be noted that the vector
fields $V_a$ in (\ref{ham-vc}) are not necessarily divergence-free,
i.e. $\partial_\mu V_a^\mu = -
\frac{\partial \Theta^{\mu\nu} (x)}{\partial x^\mu}
\frac{\partial C_a(x)}{\partial x^\nu} \neq 0$, although the vector
fields in a Darboux frame defined by (\ref{vector-v}) are
divergence-free. It should be the case as the divergence-free
condition is not covariant under general coordinate transformations.
Therefore the vector fields $V_a$ in a general coordinate system
generate longitudinal as well as transverse components
altogether.\footnote{\label{hsy}One of us (HSY) wants to confess
that he did not clearly recognize this fact before. Regrettably, in
some previous works \cite{lee-yang,hsy-ijmp09,hsy-jhep09},
Hamiltonian vector fields had been partially expressed in the
Darboux frame like (\ref{vector-v}). But, note that the form
invariance (\ref{px-bracket}) of Poisson brackets under a general
coordinate transformation corresponds to the diffeomorphism symmetry
in general relativity. In other words, one can choose a Darboux
frame with impunity to formulate the emergent gravity in the
top-down approach. (``It is impossible to study this remarkable
theory with experiencing at times the strange feeling that the
equations and formulas somehow have a proper life, that they are
smarter than we, smarter than the author himself, and that we
somehow obtain from them more than was originally put into them". --
Heinrich Hertz)} So they can be related to a basis of orthonormal
tangent vectors $E_a = E_a^\mu \partial_\mu \in \Gamma(TM)$ and
cotangent vectors (vierbeins or tetrads) $e^a = e^a_\mu dx^\mu \in
\Gamma(T^* M)$ by
\cite{ajs88,mn89,joyce,hsy-jhep09}
\begin{equation}\label{v-e}
    V_a = \lambda E_a \in \Gamma(TM), \qquad
    e^a = \lambda v^a \in \Gamma(T^* M)
\end{equation}
with $\lambda \in C^\infty(M)$ to be determined. In the next section
we will explain how to determine $\lambda$ from symplectic gauge
fields. The gravitational metric emergent from symplectic gauge
fields in (\ref{nc-c}) is thus given by
\begin{eqnarray}\label{eg-metric}
    ds^2 &=& g_{\mu\nu} dx^\mu dx^\nu = e^a \otimes e^a \nonumber \\
    &=& \lambda^2 v^a \otimes v^a = \lambda^2 v_\mu^a v_\nu^a dx^\mu
    dx^\nu.
\end{eqnarray}
As will be shown in the next section, the equations of motion for
U(1) gauge fields $\widehat{A}_\mu (y)$ over the symplectic vacuum
$\langle C_a \rangle_{\mathrm{vac}} = B_{ab} y^b$ will be
transformed to the gravitational field equations for the metric
(\ref{eg-metric}) \cite{lee-yang,review4}. This completes the idea
to construct Einstein gravity from U(1) gauge fields on a symplectic
manifold $(M, B)$, which we call top-down formulation to compare
with the bottom-up approach being the main theme of this paper.

Now we want to invert the procedure of emergent gravity to find
corresponding U(1) gauge fields on a Poisson manifold given a
Riemannian metric $(M, g)$. Suppose that a Riemannian metric $(M,
g)$ is given. One can determine $\lambda$ by solving
(\ref{gauge-phi}) and then the vector fields $V_a$ are determined by
(\ref{v-e}). After that, the vector fields $V_a$ are mapped to
symplectic gauge fields $C_a(y)$ as the system of D-module which is
characterized by (\ref{vec-comp}) and provides enough data to deduce
the equations of motion for symplectic gauge fields. We will examine
this bottom-up approach with the LeBrun metric
\cite{lebrun} that is the most general scalar-flat K\"ahler metric
with a U(1) isometry and contains the Gibbons-Hawking metric
\cite{gh-metric}, the real heaven \cite{boy-fin,bakas-sfet}
as well as the multi-blown up Burns metric which is a scalar-flat
K\"ahler metric on $\mathbb{C}^2$ with $n$ points blown up. (See
\cite{lebrun-cmp,xiao} for the Burns metric on the blow-up of
$\mathbb{C}^2$ at the origin.) The bottom-up approach clarifies some
important issues in emergent gravity, one of which was already
stated in footnote \ref{hsy}.

The paper is organized as follows. In section 2, we review how
Einstein gravity arises from the emergent metric (\ref{eg-metric}).
Especially, we explain in detail how to determine $\lambda$ in the
top-down and bottom-up approaches. In section 3, we digest the most
general scalar-flat K\"ahler metric with a U(1) isometry constructed
by LeBrun \cite{lebrun}. We will summarize essential ingredients of
the LeBrun metric for the bottom-up approach. In section 4, we
specify symplectic gauge fields obtained from the LeBrun metric and
derive the equations of motion for corresponding U(1) gauge fields.
Finally, in section 5, we conclude the bottom-up approach of
emergent gravity with several remarks and some open issues.

\section{Emergent gravity}

Let us recapitulate the underlying idea of emergent gravity. When
U(1) gauge fields have a vacuum condensate $\langle \mathcal{A}_\mu
\rangle_{\mathrm{vac}} \equiv A_\mu^{(0)}$ which admits a symplectic
structure $B= dA^{(0)}$ on the vacuum, the fluctuations of dynamical
gauge fields $A_\mu$ will be superposed on the vacuum gauge field
$A_\mu^{(0)}$ to yield $\mathcal{A}_\mu = A^{(0)}_\mu + A_\mu$.
Therefore the electromagnetic force $F=dA$ manifests itself only as
the deformation of underlying symplectic structure $B$ because the
total field strength is now given by $\mathcal{F} = d\mathcal{A} =
dA^{(0)} + dA = B + F$ and hence $d\mathcal{F} =0$. Since $F=dA$
describes a fluctuation around the vacuum $B$ and $F \to 0$ at an
asymptotic infinity, one can safely assume that $\mathcal{F}$ is
nondegenerate everywhere. Hence one can conclude that $(M,
\mathcal{F})$ defines a (dynamical) symplectic manifold. Then the
Darboux theorem in symplectic geometry implies that there always
exists a coordinate transformation on a local Darboux chart $U
\subset M$ to locally eliminate the electromagnetic force $F=dA$ on
$U$. This novel form of the equivalence principle for the
electromagnetic force implies \cite{hsy-jhep09,lee-yang,review4}
that the electromagnetism describing a dynamical symplectic manifold
$(M, \mathcal{F})$ corresponds to a geometry of spacetime manifold
$M$ whose metric is given by (\ref{eg-metric}).

In the top-down approach described above, one can calculate the
vector fields $V_a$ defined by (\ref{vector-v}) or (\ref{ham-vc})
after a general coordinate transformation only if symplectic gauge
fields $C_a(y)$ are known. However one has to know $\lambda \in
C^\infty(M)$ in order to completely determine the metric
(\ref{eg-metric}) from symplectic gauge fields in (\ref{nc-c}). We
will explain in detail how to determine $\lambda$ when the vector
fields $V_a$ are known.

First let us define the covariant divergence of inverse vierbein
$E_a^\mu$ by
\begin{eqnarray}\label{co-div}
    \nabla_\mu E^\mu_a &=& \partial_\mu E_a^\mu +
    {\Gamma_{\mu\nu}}^\mu E_a^\nu \nonumber \\
    &=& \partial_\mu E_a^\mu + E_a^\mu \partial_\mu \log
    \det e_\nu^a \nonumber \\
    &=& - \omega_{bab} \equiv - \phi_a
\end{eqnarray}
where ${\Gamma_{\mu\nu}}^\rho$ and ${{\omega_\mu}^a}_b$ are the
Levi-Civita and spin connections in general relativity,
respectively, and we used the well-known relation
${\Gamma_{\mu\nu}}^\mu = \partial_\nu \log \sqrt{\det g_{\mu\nu}} =
{{\omega_\mu}^a}_b E^\mu_a e^b_\nu + E^\mu_a \partial_\mu e_\nu^a$.
Let us introduce the structure equation for vector fields $E_a =
E_a^\mu \partial_\mu \in \Gamma(TM)$ defined by
\begin{equation}\label{estr-eq}
    [E_a, E_b] = -{f_{ab}}^c E_c
\end{equation}
where the structure coefficients are given by
\begin{equation}\label{estr-coeff}
    {f_{ab}}^c = E_a^\mu E_b^\nu (\partial_\mu e_\nu^c - \partial_\nu
    e_\mu^c).
\end{equation}
After imposing the torsion-free condition, $T^a = de^a +
{\omega^a}_b \wedge e^b = 0$, the spin connection
${{\omega_\mu}^a}_b$ can be completely determined in terms of the
structure coefficients in (\ref{estr-coeff}) as
\begin{equation}\label{spin-f}
\omega_{abc} =  E_a^\mu {\omega_\mu}_{bc} = \frac{1}{2} (f_{abc} -
f_{bca} + f_{cab}).
\end{equation}
From either (\ref{estr-coeff}) or (\ref{spin-f}), one can easily
derive the relation $\omega_{bab} = f_{bab} = \phi_a$, i.e.,
\begin{equation}\label{div-phi}
    \nabla \cdot E_a = - f_{bab} = -\phi_a.
\end{equation}

As was rigorously shown in \cite{grant} (see also
\cite{hsy-jhep09}), by performing a local SO(4) rotation of basis
vectors $E_a$, one can always achieve the gauge condition $\phi_a =
- E_a \log \lambda$ and so
\begin{equation}\label{gauge-phi}
    \nabla \cdot E_a = - \phi_a = E_a \log \lambda.
\end{equation}
This means \cite{mn89} that one can choose $\lambda$ by a local
frame rotation such that the vector field $E_a$ preserves the volume
form $\widetilde{\nu} = \lambda^{-1} \nu_g$ where $\nu_g = e^1
\wedge
\cdots \wedge e^4 = \sqrt{\det g_{\mu\nu}} d^4 x$ is the Riemannian volume form.
This can be checked as follows:
\begin{eqnarray} \label{e-preserve}
  \mathcal{L}_{E_a} \widetilde{\nu} &=& d \iota_{E_a} \Big( \lambda^{-1} \det e_\mu^a
  d^4 x \Big) \nonumber \\
  &=& d \Big( \lambda^{-1} \det e_\nu^a \sum_{\mu = 1}^4 (-1)^{\mu-1} E_a^\mu
  dx^1 \wedge \cdots \wedge \widehat{dx^\mu} \wedge \cdots \wedge
  dx^4 \Big) \nonumber \\
  &=& \Big( \partial_\mu E_a^\mu + E_a^\mu \partial_\mu \log
    \det e_\nu^a  - E_a \log \lambda \Big) \widetilde{\nu}
    \nonumber \\
    &=& \Big( \nabla \cdot E_a  - E_a \log \lambda \Big) \widetilde{\nu} = 0
\end{eqnarray}
where $\widehat{dx^\mu}$ denotes the omission of $dx^\mu$. In the
above calculation, we used the Cartan's homotopy formula
\cite{sg-book1,sg-book2}
\begin{equation}\label{lie-der}
\mathcal{L}_X = d\iota_X + \iota_X d
\end{equation}
for Lie derivative $\mathcal{L}_X$ along a vector field $X \in
\Gamma(TM)$ which is an important formula in differential geometry.
Given the relation (\ref{v-e}), the equation (\ref{e-preserve})
suggests that the vector field $V_a$ preserves the volume form $\nu
= \lambda^{-2}
\nu_g = \lambda^2 v^1 \wedge \cdots \wedge v^4$, which will be called the symplectic
volume form, due to the relation \cite{mn89}
\begin{equation}\label{lie-der-vol}
0 = \mathcal{L}_{E_a} \widetilde{\nu} = \mathcal{L}_{\lambda^{-1}
V_a} \widetilde{\nu} = \mathcal{L}_{V_a} (\lambda^{-1}
\widetilde{\nu}) = \mathcal{L}_{V_a} \nu.
\end{equation}
The above equation means (by the same calculation as
(\ref{e-preserve})) that
\begin{eqnarray} \label{v-preserve}
  \mathcal{L}_{V_a} \nu  &=& \Big( \partial_\mu V_a^\mu + V_a^\mu \partial_\mu \log
    \det v_\nu^a  + 2 V_a \log \lambda \Big) \nu \nonumber \\
    &=& \Big( \nabla \cdot V_a  + 2 V_a \log \lambda \Big) \nu
    \nonumber \\
    &=& \Big( -g_{bab}  + 2 V_a \log \lambda \Big) \nu = 0.
\end{eqnarray}
In the last step of (\ref{v-preserve}), we have introduced the
structure equation for the vector fields $V_a$ defined by
\begin{equation}\label{vstr-eq}
    [V_a, V_b] = -{g_{ab}}^c V_c.
\end{equation}

In the top-down approach, on one hand, we know $V_a \in \Gamma(TM)$
by (\ref{vector-v}) derived from symplectic gauge fields given by
(\ref{nc-c}). Thus one can solve (\ref{v-preserve}) to determine
$\lambda$ and hence determine the Riemannian metric
(\ref{eg-metric}) using the relation (\ref{v-e}). In this way, one
can completely determine the gravitational metric (\ref{eg-metric})
emergent from U(1) gauge fields. In the bottom-up approach, on the
other hand, we know a metric $(M, g)$ instead, i.e. $E_a \in
\Gamma(TM)$. Then one can solve (\ref{e-preserve}) to determine
$\lambda$ and so the vector fields $V_a \in \Gamma(TM)$ are
determined by (\ref{v-e}). When $\lambda$ is known, one can also
construct the symplectic volume form $\nu = \lambda^2 v^1 \wedge
\cdots \wedge v^4$ which leads to the relation \cite{hsy-jhep09}
\begin{equation}\label{lambda}
    \lambda^2 = \nu(V_1, \cdots, V_4).
\end{equation}
After determining $V_a$'s, one can try to solve (\ref{vec-comp}) to
yield corresponding symplectic gauge fields $C_a(y)$. In the end,
one may derive the equations of motion for the dynamical gauge
fields $\widehat{A}_\mu (y)$. We will illustrate later with some
examples how this bottom-up approach nicely works.

Note that the symplectic gauge fields in (\ref{nc-c}) are obtained
by solving the Darboux transformation (\ref{darboux-tr}) and they
completely determine the gravitational metric (\ref{eg-metric}).
We want to emphasize that the emergence of gravity originates from the global existence
of the one-parameter family of diffeomorphisms describing the local deformation of
an initial symplectic structure $B$ due to the electromagnetic force $F = dA$.
This essential point can be understood as follows \cite{hsy-jhep09}. The symplectic structure $B$ is
a nondegenerate, closed 2-form, i.e. $dB = 0$. Therefore the symplectic structure $B$
defines a bundle isomorphism $B : TM \to T^* M$ by $X \mapsto A = \iota_X B$
where $\iota_X$ is an interior product with respect to a vector field $X \in \Gamma(TM)$.
Then the electromagnetic force can be represented by $F = dA = d \iota_X B
= {\cal L}_X B$ where the formula (\ref{lie-der}) and $dB=0$ were used.
This means that the electromagnetic force $F = dA = {\cal L}_X B$ can always be
eliminated by a coordinate transformation generated by the vector field $X$.
(See eq.(23) in \cite{review4} for an explicit verification.)
This fact vindicates that the emergent gravity reproduces general relativity which
also respects diffeomorphism symmetry.

Therefore one can interpret the Darboux transformation
(\ref{darboux-tr}) in symplectic geometry from the viewpoint of
emergent gravity described by the metric (\ref{eg-metric}) (i.e., in
the context of Riemannian geometry). First one can notice that, when
fluctuations are turned off, i.e. $F(X) = 0$, the symplectic gauge
field is given by $C_a(y) = B_{ab} y^b$ and the corresponding vector
fields reduce to $V_a = \delta_a^\mu \partial_\mu$ and so $\lambda
=1$ by (\ref{lambda}). In this case, one can immediately see that
the metric (\ref{eg-metric}) becomes flat, $g_{\mu\nu} =
\delta_{\mu\nu}$, and the symplectic volume form reduces to $\nu
=d^4x = \frac{1}{2\mathrm{Pf}B} B \wedge B$ which will be called an
asymptotic volume form. Hence it turns out
\cite{lee-yang,review4,hsy-jpcs12} that the flat spacetime is emergent from the
vacuum condensate which admits an underlying symplectic structure
$B$ to the vacuum. Now, if one turns on fluctuations, i.e. $F(X)
\neq 0$, the symplectic gauge field will deviate from the vacuum one
and it is given by (\ref{nc-c}). As a result, the metric
(\ref{eg-metric}) will also deviate from the flat metric, namely,
$V_a: \delta_a^\mu
\partial_\mu \to V_a^\mu (y)
\partial_\mu$ and $ g_{\mu\nu}: \delta_{\mu\nu} \to g_{\mu\nu}(x)$.
But, according to the Darboux theorem or the Moser lemma, one can
properly choose a Darboux frame, say on $U \subset M$ that locally
nullifies the fluctuations, and the metric (\ref{eg-metric}) on the
Darboux chart $U \subset M$ then locally looks like a flat metric,
i.e. $g_{\mu\nu}|_U = \delta_{\mu\nu}$. Therefore it would be
reasonable to think that a local Darboux chart in symplectic
geometry corresponds to a local inertial frame in general
relativity.\footnote{There is a subtle but important difference
between the Riemannian and symplectic geometries \cite{hsy-jhep09}.
Strictly speaking, the equivalence principle in general relativity
is a point-wise statement (up to first-order differentials of
metric) at a given point $P$ while the Darboux theorem in symplectic
geometry is defined on an entire neighborhood around $P$. This is
the reason why there exist local invariants, e.g. curvature tensors,
in Riemannian geometry but there is no such kind of local invariant
in the symplectic geometry. This raises a question how Riemannian
geometry is emergent from symplectic geometry though their local
geometries are in sharp contrast with each other. A possible
resolution was suggested in \cite{hsy-jhep09} (see section 2.3). See
also \cite{hsy-ijmp09}.} Consequently, if Einstein gravity arises
from symplectic gauge fields in the way we have described, the
equivalence principle, the most important property in general
relativity, might be explained by the Darboux theorem and the Moser
lemma in symplectic geometry \cite{lee-yang}.

The condition (\ref{v-preserve}) says that the vector fields $\{ V_a
\}$ are volume-preserving with respect to the volume element
$\nu = \lambda^{-2} \nu_g$. Suppose that we have chosen a Darboux
frame where the ordinary divergence-free condition $\partial_\mu
V_a^\mu = 0$ is obeyed. For such a case, we have the relation
\begin{equation}\label{lambda-rel}
    \lambda^2 = \det V_a^\mu = \det e_\mu^a = \sqrt{\det
    g_{\mu\nu}}
\end{equation}
and the symplectic volume $\nu$ is equal to the asymptotic volume,
i.e.,
\begin{equation}\label{asymp-vol}
    \nu = \frac{1}{2\mathrm{Pf}B} B \wedge B = d^4 y.
\end{equation}
Therefore the symplectic volume $\nu$ remains the same as the
asymptotic volume (\ref{asymp-vol}) even after turning on the
fluctuations. Actually this is known as the Liouville theorem in
Hamiltonian mechanics, since the vector fields $\{ V_a \}$ in this
case are usual Hamiltonian vector fields. But we have remarked in
section 1 that the vector fields in a general coordinate system do
not always satisfy the condition $\partial_\mu V_a^\mu = 0$. In this
case the symplectic volume form is not equal to the asymptotic one.

Let us explore gravitational field equations for the metric
(\ref{eg-metric}) emergent from symplectic gauge fields in
(\ref{nc-c}). First note the following relations:
\begin{eqnarray} \label{cc-f}
 \{ C_a, C_b \}_\theta  &=& - B_{ab} + \partial_a \widehat{A}_b -
 \partial_b \widehat{A}_a +  \{ \widehat{A}_a, \widehat{A}_b \}_\theta \nonumber \\
 &\equiv& - B_{ab} + \widehat{F}_{ab}, \\
 \label{ccc-df}
 \{C_a, \{ C_b, C_c \}_\theta \}_\theta &=& \partial_a \widehat{F}_{bc}
  +  \{ \widehat{A}_a, \widehat{F}_{bc} \}_\theta \nonumber \\
  &\equiv& \widehat{D}_a \widehat{F}_{bc}.
\end{eqnarray}
Using the identity (\ref{px-bracket}) and the Lie algebra
homomorphism (\ref{g-lie-homo}), one can complete the important
isomorphism $\mathfrak{H}: \mathfrak{P}(M) \to \mathfrak{X}(M)$
between the set of symplectic gauge fields in (\ref{nc-c}) and the
vector fields in (\ref{ham-vc}) which is represented by
\cite{hsy-jhep09,yasi-prd10}
\begin{eqnarray} \label{hom-f}
&& X_{\widehat{F}_{ab}} = X_{ \{ C_a, C_b \}_\Theta} = [V_a, V_b], \\
\label{hom-df}
&&  X_{\widehat{D}_a \widehat{F}_{bc}} = X_{\{C_a, \{ C_b, C_c
\}_\Theta \}_\Theta} = [V_a, [V_b, V_c]].
\end{eqnarray}
Adopting the same method as (\ref{vec-comp}), one can derive from
(\ref{hom-f}) the following relation
\begin{equation}\label{vv-f}
\{\widehat{F}_{ab}, x^\mu \}_\theta (y) = - {g_{ab}}^c V_c^\mu(x)
\end{equation}
where ${g_{ab}}^c$ are structure coefficients in (\ref{vstr-eq}). In
the bottom-up approach, the right-hand side of (\ref{vv-f}) is
determined by a given metric and so, in principle, one can solve it
to determine the field strength $\widehat{F}_{ab}(y)$ of symplectic
gauge fields.

A remarkable point is that the electromagnetism on a symplectic manifold $(M,B)$
is completely described by the Poisson algebra $\mathfrak{P}(M) = (C^\infty(M),
\{-, -\}_\theta)$ \cite{review4}. For example, the action is given by
\begin{equation}\label{action-poisson}
    S = \frac{1}{4 g^2_{YM}} \int d^4 x \{ C_a, C_b \}^2_\theta.
\end{equation}
The identity (\ref{hom-df}) provides us a direct map
\cite{lee-yang,hsy-jpcs12,hsy-jhep09,yasi-prd10} to connect the
equations of motion of symplectic gauge fields derived from the action (\ref{action-poisson})
to gravitational field equations for the emergent metric (\ref{eg-metric}):
\begin{eqnarray} \label{map-jacobi}
 && \widehat{D}_a \widehat{F}_{bc} + \mathrm{cyclic}  = 0 \qquad \Leftrightarrow
 \qquad [V_a, [V_b, V_c]] + \mathrm{cyclic} = 0, \\
  \label{map-eom}
 && \widehat{D}_b \widehat{F}_{ab}  = 0 \qquad \Leftrightarrow
 \qquad [V_b, [V_a, V_b]] = 0.
\end{eqnarray}
It can be shown (even in any $2n$-dimensions)
\cite{hsy-jhep09,hsy-jpcs12} that the right-hand side of
(\ref{map-jacobi}) is precisely equivalent to the first Bianchi
identity of Riemann curvature tensors, i.e.,
\begin{equation} \label{1-bianchi}
 [V_a, [V_b, V_c]] + \mathrm{cyclic} = 0  \qquad \Leftrightarrow
 \qquad R_{[abc]d} = 0,
\end{equation}
where $[abc]$ denotes the cyclic permutation of indices. The
equations of motion (\ref{map-eom}) leads to a cryptic result for
Ricci tensors \cite{hsy-jhep09,hsy-jpcs12}
\begin{equation} \label{emergent-einstein}
  R_{ab} = - \frac{1}{\lambda^2} \Big[ g^{(+)i}_d g^{(-)j}_d
  \Big(\eta^i_{ac} \overline{\eta}^{j}_{bc}
  + \eta^i_{bc} \overline{\eta}^{j}_{ac} \Big) - g^{(+)i}_c
  g^{(-)j}_d \Big(\eta^i_{ac} \overline{\eta}^{j}_{bd}
  + \eta^i_{bc} \overline{\eta}^{j}_{ad} \Big) \Big]
\end{equation}
where $\eta^i_{ab}$ and $\overline{\eta}^{i}_{ab}$ are self-dual and
anti-self-dual 't Hooft symbols. To get the result
(\ref{emergent-einstein}), we have defined the canonical
decomposition of the structure equation (\ref{vstr-eq})
\begin{equation}\label{def-g}
g_{abc} = g^{(+)i}_c \eta^i_{ab} + g^{(-)i}_c
\overline{\eta}^{i}_{ab}.
\end{equation}

A notable point is that the right-hand side of
(\ref{emergent-einstein}) consists of purely interaction terms
between self-dual and anti-self-dual parts in (\ref{def-g}) which is
the feature withheld by matter fields only \cite{oh-yang,loy}. A
gravitational instanton which is a Ricci-flat, K\"ahler manifold can
be understood as either $g^{(-)i}_c = 0$ (self-dual) or $g^{(+)i}_c
= 0$ (anti-self-dual) in terms of (\ref{def-g}) and so $R_{ab} = 0$
in (\ref{emergent-einstein}). Hence, the result
(\ref{emergent-einstein}) is consistent with the Ricci-flatness of
gravitational instantons. However a unique property of
(\ref{emergent-einstein}) is to contain a nontrivial trace
contribution, i.e., a nonzero Ricci scalar, due to the second part
which is non-existent in Einstein gravity as was recently shown in
\cite{loy}. The content of the energy-momentum tensor defined by the
right-hand side of (\ref{emergent-einstein}) becomes manifest by
decomposing it into two parts, denoted by $8\pi G T_{ab}^{(M)}$ and
$8\pi G T_{ab}^{(L)}$, respectively \cite{hsy-jhep09,hsy-jpcs12}:
\begin{eqnarray} \label{emt-max}
8\pi G T_{ab}^{(M)} &=& - \frac{1}{\lambda^2} \Big( g_{acd} g_{bcd}
- \frac{1}{4} \delta_{ab} g_{cde} g_{cde} \Big), \\
\label{emt-liouville}
8\pi G T_{ab}^{(L)} &=&  \frac{1}{2 \lambda^2} \Big( \rho_a
\rho_b - \Psi_a \Psi_b - \frac{1}{2}\delta_{ab} \big(\rho_c^2 -
\Psi_c^2 \big) \Big),
\end{eqnarray}
where
\begin{equation}\label{rho-psi}
    \rho_a \equiv g_{bab}, \qquad \Psi_a \equiv - \frac{1}{2}
    \varepsilon^{abcd} g_{bcd}.
\end{equation}
The first energy-momentum tensor (\ref{emt-max}) is traceless, i.e.
$8\pi G T_{aa}^{(M)} = 0$, which is a consequence of the identity
$\eta^i_{ab} \overline{\eta}^{j}_{ab} = 0$ when applied to the first
part of (\ref{emergent-einstein}).  The Ricci scalar $R
\equiv R_{aa}$ can be calculated by (\ref{emt-liouville}) to
yield
\begin{equation}\label{r-scalar}
    R = \frac{1}{2 \lambda^2} \Big( \rho_a^2 - \Psi_a^2 \Big).
\end{equation}
The equation (\ref{r-scalar}) immediately leads to the conclusion
that a four-manifold emergent from pure symplectic gauge fields
(without source terms) can have a vanishing Ricci scalar if and only
if
\begin{equation}\label{scalar-dual}
    \rho_a = \pm \Psi_a
\end{equation}
that is similar to the self-duality equation. When the relation
(\ref{scalar-dual}) is obeyed, the second energy-momentum tensor
$8\pi G T_{ab}^{(L)}$ identically vanishes. In section 4 we will
show that the LeBrun metric \cite{lebrun} satisfies the relation
(\ref{scalar-dual}) and so it can arise in emergent gravity from
pure symplectic gauge fields.

It would be worthwhile to remark that a four-manifold with a
vanishing Ricci scalar cannot be realized as a vacuum solution of
Einstein gravity without matter fields. Indeed the Einstein's
equation can be written as $R_{\mu\nu} - \frac{1}{4} g_{\mu\nu} R =
8 \pi G \widetilde{T}_{\mu\nu}$ where $\widetilde{T}_{\mu\nu} =
T_{\mu\nu} - \frac{1}{4} g_{\mu\nu} T$ is a traceless
energy-momentum tensor which annuls a possible cosmological
constant. For a scalar-flat four-manifold, the Einstein equations
reduce to $R_{\mu\nu} = 8 \pi G \widetilde{T}_{\mu\nu}$. This means,
a scalar-flat manifold can only arise from a traceless (conformal)
matter source. This condition can be realized in Einstein-Yang-Mills
systems in four-dimensions. For instance, the LeBrun metric is a
solution of Euclidean Einstein-Maxwell theory
\cite{lebrun-hitchin,bnw11,bnw12}. There are some reasons that the energy-momentum tensor
(\ref{emt-max}) can be mapped to that of the usual Maxwell theory in
commutative spacetime. Indeed it was argued in
\cite{hsy-jhep09} that it can be done by reversing the map
(\ref{gierbein}). Hence, the emergent gravity shows that such a
scalar-flat four-manifold can emerge from pure Maxwell theory on a
symplectic manifold $(M, B)$.

\section{Scalar-flat K\"ahler metrics}

LeBrun found in \cite{lebrun} the explicit local form of all
Euclidean, four-dimensional K\"ahler metrics that have a U(1)
isometry and a vanishing Ricci scalar. It is then shown in
\cite{lebrun-hitchin} that these metrics are necessarily solutions
of Einstein-Maxwell theory whose electromagnetic field is related to
the K\"ahler form. The LeBrun metric takes the form
\begin{equation}\label{lebrun}
    ds^2 = w^{-1} (d\tau + A)^2 + w \big( e^u (dx^2 + dy^2) + dz^2
    \big)
\end{equation}
where $w > 0$ and $u$ are smooth real-valued functions on an open
set $U \subset \mathbb{R}^3$ which satisfy the $su(\infty)$ Toda
equation and its linearized form:
\begin{eqnarray} \label{toda}
&& \partial_x^2 u + \partial_y^2 u + \partial_z^2 (e^u) = 0,  \\
\label{l-toda}
&&  \partial_x^2 w + \partial_y^2 w + \partial_z^2 (e^u w) =0.
\end{eqnarray}
The one-form, $A$, obeys
\begin{equation}\label{1-form-a}
    dA = \partial_x w dy \wedge dz + \partial_y w dz \wedge dx
    + \partial_z (e^u w) dx \wedge dy
\end{equation}
and the closedness of $dA$, i.e. $d^2 A = 0$, is equivalent to the
equation (\ref{l-toda}). The K\"ahler form is given by
\begin{equation}\label{j-kahler}
    \Omega = (d\tau + A) \wedge dz - w e^u dx \wedge dy
\end{equation}
and the metric (\ref{lebrun}) is K\"ahler, i.e., $d\Omega = 0$.

The LeBrun metric (\ref{lebrun}) is defined by two functions,
$u(\mathbf{x})$ and $w(\mathbf{x})$, on an open set $ U
\subset \mathbb{R}^3 \ni \mathbf{x}$, satisfying (\ref{toda}) and (\ref{l-toda}),
respectively. But one may consider the function $w(\mathbf{x})$  as
a linear perturbation of $u(\mathbf{x})$ from a Toda point
$u_t(\mathbf{x})$ which satisfies the $su(\infty)$ Toda equation
(\ref{toda}). Then the equation (\ref{l-toda}) implies that a linear
deviation of the function $u(\mathbf{x})$ from a Toda point
$u_t(\mathbf{x})$, that is $u(\mathbf{x}) = u_t(\mathbf{x}) +
w(\mathbf{x})$ and so $e^{u(\mathbf{x})} \approx e^{u_t(\mathbf{x})}
+ e^{u_t(\mathbf{x})} w(\mathbf{x})$, is still a solution of the
$su(\infty)$ Toda equation (\ref{toda}). For example, if the Toda
point is $u_t = 0$, we get the Gibbons-Hawking metric and, if
$w(\mathbf{x})$ is generated by a $z$-translation from a Toda point
$u_t$ with $A_3 = 0$, i.e. $u(z + \epsilon)
\approx u_t(z) +
\epsilon \partial_z u(z) := u_t(\mathbf{x}) + w(\mathbf{x})$,
the  metric (\ref{lebrun}) gives rise to the real heaven. Indeed the
Gibbons-Hawking metric \cite{gh-metric} takes the form
\begin{equation}\label{gh-metric}
    ds^2 = w^{-1} (d\tau + A)^2 + w \big(dx^2 + dy^2 + dz^2
    \big)
\end{equation}
and $w(\mathbf{x})$ is a harmonic function on $\mathbb{R}^3$. The
U(1) gauge field $A$ satisfies the self-duality equation
\begin{equation}\label{gh-sde}
    \nabla \times \mathbf{A} = \nabla w
\end{equation}
which is precisely (\ref{1-form-a}) and is consistent with the
harmonic equation (\ref{l-toda}) (with $u=0$). And the explicit form
of the real heaven metric \cite{boy-fin,bakas-sfet} is given by
\begin{equation}\label{real-heaven}
    ds^2 = (\partial_z u)^{-1} (d\tau + a)^2 + \partial_z u \big( e^u (dx^2 + dy^2) + dz^2
    \big)
\end{equation}
where $ a = \partial_y u dx - \partial_x u dy$. Since the function
$u(\mathbf{x})$ obeys the continual Toda equation (\ref{toda}), it
is easy to see that the U(1) field strength $F = da$ is equal to
(\ref{1-form-a}).

Another interesting Toda point is given by
\begin{equation}\label{burn}
    u = \log 2z
\end{equation}
which is definitely a solution of the equation (\ref{toda}). In this
case, the so-called LeBrun-Burns metric \cite{lebrun} can be written
in the form
\begin{equation}\label{burns-metric}
    ds^2 = \zeta^2 \left( V^{-1} (d\tau + A)^2 + V \Big(\frac{dx^2 + dy^2
    + d\zeta^2}{\zeta^2} \Big) \right)
\end{equation}
by introducing a new coordinate $\zeta \equiv \sqrt{2z}$ and a new
potential $V \equiv  w e^{u} = \zeta^2 w$. Note that the
three-dimensional metric is the standard constant-curvature metric
on the hyperbolic plane $\mathbb{H}_3$:
\begin{equation}\label{ads3}
    ds^2_{\mathbb{H}_3} = \frac{dx^2 + dy^2 + d\zeta^2}{\zeta^2}.
\end{equation}
Then the equations (\ref{l-toda}) and (\ref{1-form-a}) imply that
$V$ is a harmonic function on the hyperbolic plane and $A$ satisfies
an appropriate self-duality equation on $\mathbb{H}_3$
\cite{bnw11,bnw12}:
\begin{equation}\label{h-sde}
\nabla^2_{\mathbb{H}_3} V = 0, \qquad dA = *_{\mathbb{H}_3} dV.
\end{equation}
Therefore the LeBrun-Burns metric (\ref{burns-metric}) provides a
hyperbolic analogue of the Gibbons-Hawking metric (\ref{gh-metric})
although it is not a hyper-K\"ahler manifold but just a K\"ahler
manifold with vanishing scalar curvature.

It is convenient to introduce coframes for the LeBrun metric
(\ref{lebrun})
\begin{equation}\label{coframe}
    e^1 = w^{\frac{1}{2}} e^{\frac{u}{2}} dx, \quad
    e^2 = w^{\frac{1}{2}} e^{\frac{u}{2}} dy, \quad
    e^3 = w^{\frac{1}{2}} dz, \quad
    e^4 = w^{-\frac{1}{2}} (d\tau + A),
\end{equation}
and frames
\begin{eqnarray} \label{frame}
\begin{array}{ll}
E_1 = w^{-\frac{1}{2}} e^{-\frac{u}{2}}
    \Big(\frac{\partial}{ \partial x} - A_1 \frac{\partial}{ \partial \tau} \Big), \quad &
    E_2 = w^{-\frac{1}{2}} e^{-\frac{u}{2}}
    \Big(\frac{\partial}{ \partial y} - A_2 \frac{\partial}{ \partial \tau} \Big), \\
E_3 = w^{-\frac{1}{2}} \Big(\frac{\partial}{ \partial z} - A_3
\frac{\partial}{ \partial \tau} \Big), &
E_4 = w^{\frac{1}{2}} \frac{\partial}{ \partial \tau}.
\end{array}
\end{eqnarray}
The Poisson bivector $\Theta \equiv \Omega^{-1}
\in \Gamma (\wedge^2 TM)$ determined by the K\"ahler form $\Omega = - (e^1
\wedge e^2 + e^3 \wedge e^4)$ takes the form
\begin{equation}\label{lebrun-poisson}
\Theta = \frac{1}{2} \Theta^{\mu\nu} (x)
\frac{\partial}{\partial x^\mu} \bigwedge \frac{\partial}{\partial x^\nu}
= E_1 \wedge E_2 + E_3 \wedge E_4.
\end{equation}
For our later purpose, we present explicit forms of the K\"ahler
form and the Poisson tensor:
\begin{eqnarray} \label{matrix-k}
  && \Omega_{\mu\nu} = \left(
       \begin{array}{cccc}
         0 & - w e^{u} & A_1 & 0 \\
         w e^{u} & 0 & A_2 & 0 \\
         -A_1 & -A_2 & 0 & -1 \\
         0 & 0 & 1 & 0 \\
       \end{array}
     \right), \\
   \label{matrix-p}
  &&  \Theta^{\mu\nu} = \left(
      \begin{array}{cccc}
         0 & w^{-1} e^{-u} & 0 & -w^{-1} e^{-u} A_2 \\
         - w^{-1} e^{-u} & 0 & 0 & w^{-1} e^{-u} A_1 \\
         0 & 0 & 0 & 1 \\
         w^{-1} e^{-u} A_2 & -w^{-1} e^{-u} A_1 & -1 & 0 \\
       \end{array}
     \right).
\end{eqnarray}

It was shown in \cite{lebrun-hitchin,bnw11} that the LeBrun metric
is a solution of Euclidean Einstein-Maxwell equations
\begin{equation}\label{einstein-maxwell}
    R_{\mu\nu} = \frac{1}{2} \Big( \mathcal{F}_{\mu\rho}
    {\mathcal{F}_\nu}^\rho - \frac{1}{4} g_{\mu\nu} \mathcal{F}_{\rho\sigma}
    \mathcal{F}^{\rho\sigma} \Big).
\end{equation}
The Maxwell field strength $\mathcal{F} \equiv F + \Omega$ is given
by
\begin{equation}\label{maxwell-f}
    \mathcal{F} = - \frac{1}{2} \sum_{i=1}^3 \partial_i \Big( \frac{\partial_z u}{w} \Big)
\Omega_-^{(i)} + \Omega
\end{equation}
where $\Omega$ is the K\"ahler form (\ref{j-kahler}) and
$\Omega_-^{(i)}$ are anti-self-dual forms defined by
\begin{equation} \label{asd-form}
\begin{array}{l}
\Omega_-^{(1)} = e^{-\frac{u}{2}} (e^2 \wedge e^3 - e^1 \wedge e^4), \\
\Omega_-^{(2)} = e^{-\frac{u}{2}} (e^3 \wedge e^1 - e^2 \wedge e^4), \\
\Omega_-^{(3)} =  e^1 \wedge e^2 - e^3 \wedge e^4.
\end{array}
\end{equation}
Our convention for the anti-self-dual forms in (\ref{asd-form}) is
actually the orientation flip of the self-dual forms in
\cite{bnw11} because the Riemannian volume form in our case is given by $\nu_g =
\Upsilon \, dx \wedge dy \wedge dz \wedge d\tau$ with $\Upsilon = w e^{u}$ while
the volume form in \cite{bnw11} is given by $\nu_g =
\Upsilon \, d\tau \wedge dx \wedge dy \wedge dz$. The two-form $F = dC$
in (\ref{maxwell-f}) has a vector potential given by
\begin{equation}\label{vector-b}
C = \frac{1}{2} \left( \Big( \frac{\partial_z u}{w} \Big) (d\tau +
A) - \partial_y u dx + \partial_x u dy \right).
\end{equation}
Note that $F = dC$ identically vanishes for self-dual manifolds such
as the Gibbons-Hawking metric and the real heaven. In this case, the
Maxwell field strength $\mathcal{F} = \Omega$ is simply given by the
self-dual K\"ahler form (\ref{j-kahler}) and so the energy-momentum
tensor in (\ref{einstein-maxwell}) is identically zero to yield
Ricci-flat manifolds.

\section{U(1) gauge fields from scalar-flat K\"ahler metrics}

Now we will explore symplectic gauge fields defined by the map
(\ref{vec-comp}). First let us start with warmup examples -
gravitational instantons \cite{egh-report,opy} (hyper-K\"ahler
manifolds) in (\ref{gh-metric}) and (\ref{real-heaven}) and then
consider the general case described by the LeBrun metric
(\ref{lebrun}).

\subsection{Gibbons-Hawking metric}

For the Gibbons-Hawking metric (\ref{gh-metric}), it is easy to
solve (\ref{e-preserve}) using the inverse vierbeins in
(\ref{frame}) (with $u=0$) to determine $\lambda$ given by
\begin{equation}\label{gh-lambda}
    \lambda = w^{\frac{1}{2}}.
\end{equation}
Then the relation (\ref{v-e}) determines the vector fields $V_a =
\lambda E_a \in \Gamma(TM)$ to be \cite{joyce}
\begin{eqnarray} \label{gh-frame}
\begin{array}{ll}
V_1 = \frac{\partial}{\partial x} - A_1 \frac{\partial}{ \partial
\tau}, \quad &
    V_2 = \frac{\partial}{ \partial y} - A_2 \frac{\partial}{ \partial \tau}, \\
V_3 = \frac{\partial}{ \partial z} - A_3
\frac{\partial}{ \partial \tau}, &
V_4 = w \frac{\partial}{ \partial \tau}.
\end{array}
\end{eqnarray}
It is easy to check using (\ref{gh-sde}) that the vector fields
(\ref{gh-frame}) satisfy the anti-self-duality equation
\cite{ajs88,mn89,joyce}
\begin{equation}\label{gh-vec-sde}
    [V_a, V_b] = - \frac{1}{2} {\varepsilon_{ab}}^{cd} [V_c, V_d].
\end{equation}
Moreover the vector fields (\ref{gh-frame}) are divergence-free,
i.e., $\partial_\mu V_a^\mu (x) = 0$.

From the vector fields in (\ref{gh-frame}), we can determine the
Poisson system for symplectic gauge fields to obey:
\begin{eqnarray} \label{gh-poisson1}
&&  V_a (x^i) = \{ C_a, x^i \}_\theta (y) = \delta^i_a,  \\
\label{gh-poisson2}
&&  V_a (\tau) = \{ C_a, \tau \}_\theta (y) \equiv Y_a(x) = (-A_i,
w)(x).
\end{eqnarray}
One might try to directly solve the Poisson system
(\ref{gh-poisson1}) and (\ref{gh-poisson2}) of the partial
differential equations to determine the symplectic gauge fields
$C_a(y)$. However it is not possible unless $x^\mu (y)$ are
explicitly known. One way to avoid this is by going to a particular
frame using the symmetry (\ref{px-bracket}) where the underlying
symplectic structure is defined by the K\"ahler form
(\ref{j-kahler}) itself. In this K\"ahler frame, the vector fields
are given by
\begin{equation}\label{k-frame}
    V_a^\mu (x) = \{ C_a(x), x^\mu \}_\Theta
    = - \Theta^{\mu\nu}(x) \frac{\partial C_a(x)}{\partial
    x^\nu}
\end{equation}
using the Poisson tensor $\Theta^{\mu\nu}$ in (\ref{matrix-p}) (with
$u=0$). It will be useful to have the explicit expression for
$J_{ab}
\equiv\frac{\partial C_a(x)}{\partial x^b}$:
\begin{equation}\label{matrix-j}
    J_{ab} = \left(
                   \begin{array}{cccc}
                     0 & -w & 0 & 0 \\
                     w & 0 & 0 & 0 \\
                     -A_1 & -A_2 & -A_3 & -1 \\
                     0 & 0 & w & 0 \\
                   \end{array}
                 \right)
\end{equation}
where $a$ is the row index and $b$ is the column index. Then it is
easy to see that $J_{ab} - J_{ba} = \Omega_{ab} - w \eta^3_{ab}$
where $\Omega_{ab}$ are components of the K\"ahler form in
(\ref{j-kahler}) and $\eta^3_{ab}$ is the self-dual 't Hooft symbol.
It implies that $J_{ab} dx^a \wedge dx^b + \frac{w}{2}
\eta^3_{ab} dx^a \wedge dx^b = \Omega$ is a closed two-form.

In order to get a better handle over the differential equation
(\ref{matrix-j}), it would be worthwhile to appreciate that the
symplectic gauge fields $C_a(x)$ in (\ref{nc-c}) are non-local
functions in general, effectively describing the dynamics of
dipole-like objects. Actually one should not insist that $C_a(x)$
are local functions because $\widehat{A}_a(x)$ in (\ref{nc-c})
corresponds to a leading approximation of NC U(1) gauge fields up to
$\mathcal{O}(\theta)$ whose physical excitations are described by NC
dipoles--weakly interacting, nonlocal objects
\cite{biga-suss,sjrey}. In order to clarify this point, let us
introduce an open Wilson line
\cite{owl-1,owl-2,owl-3} which plays a crucial role in the
Seiberg-Witten map \cite{liu}. First consider a path $P$
parameterized by $\zeta^\mu (\sigma) =
\theta^{\mu\nu} k_\nu \sigma$ with $0 \le \sigma \le 1$ and define a curve by
\begin{equation}\label{curve}
    x^\mu(\sigma) = x_0^\mu + \zeta^\mu(\sigma)
\end{equation}
with $x^\mu(\sigma = 0) \equiv x_0^\mu$ and $x^\mu(\sigma = 1)
\equiv x^\mu$. If one considers the following symplectic gauge
fields defined by (see the second equation of (54) in \cite{okoo})
\begin{equation}\label{symp-gauge}
    \int_0^1 d\sigma \frac{dx^\lambda(\sigma)}{d\sigma}
    J_{\mu\lambda} \big( x(\sigma) \big) = C_\mu(x) -
    C_\mu(x_0),
\end{equation}
they obey (\ref{matrix-j}). Here we are applying the following
formula
\begin{equation}\label{owl-formula}
\frac{\partial}{\partial x^\mu} \int_0^1 d\sigma
     \frac{dx^\lambda(\sigma)}{d\sigma}
    K \big( x(\sigma) \big) = \delta^\lambda_\mu K(x)
\end{equation}
for some differentiable function $K(x)$. Note that the dipole field
in (\ref{symp-gauge}) is an extended object with size $|\zeta| =
|x-x_0|$ but the vector fields $V_a$ become local as usual although
symplectic gauge fields could be
non-local.\footnote{\label{gauge-trans}It should be noticed that the
symplectomorphism, $x^\mu = y^\mu + \{y^\mu, \phi\}_\theta$, is in
fact equivalent to U(1) gauge transformation \cite{hsy-ijmp09}. In
particular the symplectomorphism with the gauge parameter $\phi =
k_\mu y^\mu$ generates a translation $x^\mu = y^\mu + \zeta^\mu$
with $\zeta^\mu = \theta^{\mu\nu} k_\nu$. Hence the two points
$x^\mu$ and $y^\mu$ are on the same gauge orbit, i.e. $x^\mu \sim
y^\mu$. Therefore the dipole field (\ref{symp-gauge}) actually
behaves like a closed loop in ``physical phase space." This closed
string picture for dipole fields was further elaborated in
\cite{sjrey}.}

It seems quite nontrivial to solve (\ref{symp-gauge}) for general
(multi-centered) Gibbons-Hawking metric (even for the simplest
Eguchi-Hanson metric demands for a separate work
\cite{our-future}). Rather we will determine the equations of motion
that the symplectic gauge fields must satisfy. In this respect, we
can apply the Lie algebra homomorphism (\ref{hom-f}) to the
anti-self-duality equation (\ref{gh-vec-sde}) to show that the U(1)
field strength is anti-self-dual, i.e.,
\begin{equation}\label{ghf-sde}
\widehat{F}_{ab} = - \frac{1}{2} {\varepsilon_{ab}}^{cd}
\widehat{F}_{cd}.
\end{equation}
This is consistent with the result \cite{hsy-epl09} in the top-down
approach that the Gibbons-Hawking metric arises from symplectic U(1)
instantons. Furthermore, since the vector fields in (\ref{gh-frame})
arise from a specific solution, the symplectic gauge fields for the
Gibbons-Hawking metric (\ref{gh-metric}) are further constrained. In
order to discuss this aspect, it is convenient to introduce the
Jacobiator defined by
\begin{equation}\label{jacobiator}
    J(f,g,h) \equiv \{\{ f, g \}_\theta, h \}_\theta + \{ \{ g, h \}_\theta, f \}_\theta
    + \{ \{ h, f \}_\theta, g \}_\theta
\end{equation}
for $f,g,h \in C^\infty (M)$. The Jacobi identity $J(C_a,C_b, x^i) =
0$ then leads to the result $\{ \widehat{F}_{ab}, x^i \}_\theta =
0$. Combining it with the Lie algebra relation (\ref{vv-f}) yields
the condition
\begin{equation}\label{gh-lie1}
\{ \widehat{F}_{ab}, x^i \}_\theta = - {g_{ab}}^i = 0.
\end{equation}
It is easy to check that the vector fields in (\ref{gh-frame})
indeed satisfy (\ref{gh-lie1}) and nonzero components are given by
\begin{equation}\label{str-coeff}
 {g_{ab}}^4 = - \frac{1}{w} (\delta^i_a \partial_i Y_b - \delta^i_b
 \partial_i Y_a).
\end{equation}
Similarly the Jacobi identity $J(C_a,C_b, \tau) = 0$ leads to the
relation
\begin{equation}\label{gh-lie2}
\{ \widehat{F}_{ab}, \tau \}_\theta = - {g_{ab}}^4 w
= \{ C_a, Y_b \}_\theta - \{C_b, Y_a \}_\theta
\end{equation}
where we imposed the condition (\ref{gh-lie1}). Of course, the above
equation must be anti-self-dual with respect to $(a,b)$ index pair.

Using the same strategy as (\ref{k-frame}), one can represent
(\ref{gh-lie1}) and (\ref{gh-lie2}) in the K\"ahler frame
(\ref{matrix-p}) and the result can be written as
\begin{eqnarray} \label{gh-f-cond1}
  && \frac{\partial \widehat{F}_{ab}}{\partial x}
  = \frac{\partial \widehat{F}_{ab}}{\partial y}
  = \frac{\partial \widehat{F}_{ab}}{\partial \tau} = 0, \\
  \label{gh-f-cond2}
  && \frac{\partial \widehat{F}_{ab}}{\partial z} (x) = - {g_{ab}}^4
  w(x) = V_a(Y_b) (x) - V_b(Y_a) (x)
\end{eqnarray}
where
\begin{equation}\label{g-coeff}
{g_{ij}}^4 = \varepsilon_{ijk} \partial_k \log w,
\qquad {g_{4i}}^4 = \partial_i \log w.
\end{equation}
The above equations imply that, if we solve the self-duality
equation (\ref{ghf-sde}) with the U(1) field strength given by
\begin{equation}\label{gh-fk-frame}
\widehat{F}_{ab} (x) = B_{ab} +
\Theta^{\mu\nu}(x)
\frac{\partial C_a(x)}{\partial x^\mu}
\frac{\partial C_b(x)}{\partial x^\nu},
\end{equation}
then the dipole field for the U(1) field strength extends along
$z$-direction only (according to the formula (\ref{owl-formula})).
It will be interesting to explicitly solve (\ref{gh-f-cond2}) using
the Gibbons-Hawking metric (\ref{gh-metric}). We want to postpone
this project to future works which will be initiated in
\cite{our-future}. Anyway the bottom-up approach again proves the
equivalence \cite{sty-06,prl-06,hsy-epl09} between gravitational
instantons and symplectic U(1) instantons, rigorously established
from the top-down approach \cite{hsy-jhep09,lee-yang,hsy-jpcs12}.

\subsection{Real heaven}

The real heaven metric (\ref{real-heaven}) can be analyzed precisely
in the same way as the Gibbons-Hawking case except for the fact that
a frame rotation is necessary to solve (\ref{e-preserve}). Let us
take a particular SO(4) rotation
\begin{equation}\label{so4-rot}
    \left(
      \begin{array}{c}
        E'_3 \\
      E'_4 \\
      \end{array}
    \right) = \left(
                \begin{array}{cc}
                  \cos \frac{\tau}{2} & - \sin \frac{\tau}{2} \\
                  \sin \frac{\tau}{2} & \cos \frac{\tau}{2} \\
                \end{array}
              \right) \left(
      \begin{array}{c}
        E_3 \\
      E_4 \\
      \end{array}
    \right),
\end{equation}
leaving (1-2)-plane unchanged. In this rotated frame, it is easy to
solve (\ref{e-preserve}) using the inverse vierbeins in
(\ref{frame}) (with $w = \partial_z u$ and $A_3=0$) to determine
$\lambda$ given by
\begin{equation}\label{rh-lambda}
    \lambda = w^{\frac{1}{2}} e^{\frac{u}{2}}.
\end{equation}
The vector fields $V_a \in \Gamma(TM)$ in the rotated frame are
determined by the relation (\ref{v-e}) as (after dropping the prime)
\cite{japan-osaka}
\begin{eqnarray} \label{rh-frame}
\begin{array}{ll}
V_1 = \frac{\partial}{\partial x} - a_1 \frac{\partial}{ \partial
\tau}, \quad &
    V_2 = \frac{\partial}{ \partial y} - a_2 \frac{\partial}{ \partial \tau}, \\
V_3 = e^{\frac{u}{2}} \big( \cos \frac{\tau}{2} \frac{\partial}{
\partial z} - \partial_z u \sin \frac{\tau}{2} \frac{\partial}{
\partial \tau} \big), \quad & V_4 = e^{\frac{u}{2}} \big( \sin \frac{\tau}{2}
\frac{\partial}{ \partial z} + \partial_z u \cos \frac{\tau}{2} \frac{\partial}{
\partial \tau} \big)
\end{array}
\end{eqnarray}
where $a_i = \varepsilon_{ij} \partial_j u \;(i,j=1,2)$. The above
vector fields together with (\ref{toda}) and (\ref{1-form-a})
immediately show that they also satisfy the anti-self-duality
equation \cite{ajs88,mn89,joyce}
\begin{equation}\label{rh-vec-sde}
    [V_a, V_b] = - \frac{1}{2} {\varepsilon_{ab}}^{cd} [V_c, V_d].
\end{equation}
Furthermore the vector fields (\ref{rh-frame}) are also
divergence-free, i.e., $\partial_\mu V_a^\mu (x) = 0$. Thus the U(1)
field strength derived from the real heaven metric
(\ref{real-heaven}) must be anti-self-dual, i.e.,
\begin{equation}\label{rhf-sde}
\widehat{F}_{ab} = - \frac{1}{2} {\varepsilon_{ab}}^{cd}
\widehat{F}_{cd}.
\end{equation}
This is also consistent with the top-down approach as was shown in
\cite{hsy-epl09}.

For convenience let us explicitly rewrite the components of the
vector fields (\ref{rh-frame}) defined by (\ref{k-frame})
\begin{equation}\label{matrix-rhv}
    V_a^\mu(x) = \left(
                   \begin{array}{cccc}
                     1 & 0 & 0 & -a_1 \\
                     0 & 1 & 0 & -a_2 \\
                     0 & 0 & e^{\frac{u}{2}} \cos \frac{\tau}{2}
                     & -e^{\frac{u}{2}} \partial_z u \sin \frac{\tau}{2} \\
                     0 & 0 & e^{\frac{u}{2}}\sin \frac{\tau}{2}
                     & e^{\frac{u}{2}} \partial_z u \cos \frac{\tau}{2} \\
                   \end{array}
                 \right).
\end{equation}
The corresponding matrix $J_{ab} = \frac{\partial C_a}{\partial x^b}
= V_a^\lambda \Omega_{\lambda b}$ for the real heaven metric is
given by
\begin{equation}\label{matrix-rhj}
    J_{ab}(x) = \left(
                   \begin{array}{cccc}
                     0 & -w e^u & 0 & 0 \\
                     w e^u & 0 & 0 & 0 \\
                     - a_1 e^{\frac{u}{2}} \cos \frac{\tau}{2}
                     & -a_2 e^{\frac{u}{2}} \cos \frac{\tau}{2}
                     & -e^{\frac{u}{2}} \partial_z u \sin \frac{\tau}{2}
                     & - e^{\frac{u}{2}} \cos \frac{\tau}{2} \\
                     - a_1 e^{\frac{u}{2}} \sin \frac{\tau}{2}
                     & - a_2 e^{\frac{u}{2}} \sin \frac{\tau}{2}
                     & e^{\frac{u}{2}} \partial_z u \cos \frac{\tau}{2}
                     & - e^{\frac{u}{2}}\sin \frac{\tau}{2} \\
                   \end{array}
                 \right).
\end{equation}
The symplectic gauge fields $C_a(x)$ for the real heaven can also be
solved by introducing a dipole field similar to (\ref{symp-gauge}).
However this case is more complicated than (\ref{symp-gauge})
because the path $P$ for the open Wilson line has to be placed in
four-dimensional space parameterized by $(\tau,
\mathbf{x})$ while for the Gibbons-Hawking case it was enough to span
three-dimensional space $\mathbb{R}^3$ parameterized by
$\mathbf{x}$. Moreover the Jacobi identity $J(C_a, C_b, x^i) =
\{\widehat{F}_{ab}(x), x^i \}_\Theta = 0$ for $i=1,2$ implies only
the condition
\begin{equation}\label{rh-holo}
    \big( \partial_v - a_v \partial_\tau) \widehat{F}_{ab}(x) = 0
\end{equation}
where $v = \frac{1}{2} (x+iy), \; \partial_v = \partial_x - i
\partial_y$ and $a_v = a_1 - i a_2 = i \partial_v u$.

But we may simplify the problem as was noticed in
\cite{osaka-jmp97}. The starting point is to observe that the system
of vector fields in (\ref{rh-frame}) can be regarded as Hamiltonian
vector fields on $\mathbb{R}^2 \times \Sigma$ where $(\Sigma,
\omega)$ is a two-dimensional symplectic manifold. So let us
represent them as
\begin{equation}\label{rh-v-w}
    V_1 = \frac{\partial}{\partial x} + W_{\psi_1}, \quad
    V_2 = \frac{\partial}{\partial y} + W_{\psi_2},  \quad
    V_3 = W_{\psi_3}, \quad  V_4 = W_{\psi_4}
\end{equation}
where $W_{\psi_a} \equiv \psi_a^\alpha(x) \partial_\alpha \; (\alpha
= 1,2 \in (z, \tau))$ are Hamiltonian vector fields on $\Sigma$
associated with some functions $\psi_a \in C^\infty(\mathbb{R}^2
\times \Sigma)$. Then the anti-self-duality equation
(\ref{rh-vec-sde}) reads as \cite{osaka-jmp97}
\begin{equation}\label{rh-22sde}
 \begin{array}{l}
   \frac{\partial \psi_2}{\partial x} - \frac{\partial \psi_1}{\partial y}
   + \{\psi_1, \psi_2 \}  + \{\psi_3, \psi_4 \} = 0, \\
   \frac{\partial \psi_3}{\partial x} - \frac{\partial \psi_4}{\partial y}
   + \{\psi_1, \psi_3 \}  - \{\psi_2, \psi_4 \} = 0,  \\
   \frac{\partial \psi_4}{\partial x} + \frac{\partial \psi_3}{\partial y}
   + \{\psi_1, \psi_4 \}  + \{\psi_2, \psi_3 \} = 0,
 \end{array}
\end{equation}
where $\{\psi_a, \psi_b \} = \partial_z \psi_a \partial_\tau \psi_b
- \partial_\tau \psi_a \partial_z \psi_b$ denotes the Poisson
bracket on $\Sigma$. If we wish, $\Sigma$ can be taken as a Riemann
surface of genus $g$.

Now we note that the vector fields in (\ref{rh-v-w}) are precisely
the same as those arising from four-dimensional noncommutative U(1)
gauge theory on $\mathbb{R}_C^2 \times
\mathbb{R}_{NC}^2$ which is mapped to two-dimensional U(N $\to \infty$) gauge
theory with two adjoint scalar fields (see eq.(3.30) in
\cite{hsy-jhep09}). From the viewpoint of U(N) gauge
theory, we make the following identification:
\begin{equation}\label{2-4-id}
    \psi_i = \widehat{a}_i, \qquad \psi_3 + i \psi_4 = \Phi,
    \qquad \psi_3 - i \psi_4 = \Phi^\dagger
\end{equation}
where $\widehat{a}_i \;(i=1,2)$ and $\Phi$ are two-dimensional
U(N$\to\infty$) gauge fields and a complex adjoint scalar field on
$\mathbb{R}^2$ or equivalently four-dimensional symplectic U(1)
gauge fields on $\mathbb{R}^2
\times \Sigma$. Using the notation of (\ref{2-4-id}), the
anti-self-duality equation (\ref{rh-22sde}) can be written as (see
eq.(4.1) in
\cite{hsy-epjc09})
\begin{equation}\label{rh-bps}
    \widehat{F}_{12} = \frac{i}{2} \{ \Phi^\dagger, \Phi \}, \qquad
    D_{\bar{v}} \Phi = 0
\end{equation}
where $\widehat{F}_{12} = \partial_1 \widehat{a}_2 - \partial_2
\widehat{a}_1 + \{\widehat{a}_1, \widehat{a}_2 \}$ and $D_{\bar{v}} = D_x + i D_y$.
It is remarkable that the self-dual system (\ref{rh-vec-sde}) for
the real heaven metric reduces to the BPS equations (\ref{rh-bps})
with gauge group $G = SDiff(\Sigma)$ -- area preserving
diffeomorphisms on a Riemann surface $\Sigma$, for example, or U(N
$\to \infty$) after the quantization of $(\Sigma,
\omega)$. Furthermore it was shown in \cite{hsy-epjc09} that the
BPS equations (\ref{rh-bps}) can be recast into the equation of
motion derived from the two-dimensional U(N) chiral model governed
by the action
\begin{equation}\label{2d-chiral}
    S = \frac{1}{2} \int d^2 z \mathrm{Tr} \partial_\mu h^{-1}
    \partial_\nu h \delta^{\mu\nu}
\end{equation}
where a group element $h(z)$ defines a map from $\mathbb{R}^2$ to
$GL(N, \mathbb{C})$ group, which is contractible to $U(N)
\subset GL(N, \mathbb{C})$. It has been known \cite{qhpark,ward,husain} that
the chiral model (\ref{2d-chiral}) in the $N \to \infty$ limit
describes a self-dual spacetime whose equations of motion take the
Pleba\'nski form of self-dual Einstein equations \cite{plebanski}.

Finally it is not difficult to solve the coupled equations
(\ref{rh-22sde}) to determine the symplectic gauge fields in
(\ref{2-4-id}) and the result is already known thanks to
\cite{osaka-jmp97}:
\begin{equation}\label{rh-gsol}
\widehat{a}_1 (x) = - \int^z \frac{\partial u}{\partial y} dz, \qquad
\widehat{a}_2 (x) = \int^z \frac{\partial u}{\partial x} dz, \qquad
\Phi (x) = 2 e^{\frac{u + i\tau}{2}}.
\end{equation}
Therefore we have got the solution (\ref{rh-gsol}) of the BPS
equations (\ref{rh-bps}) based on the bottom-up
approach.\footnote{Unfortunately we cannot make a similar reduction
for the Gibbons-Hawking metric. The system (\ref{gh-frame}) consists
of vector fields on $\mathbb{R}^3
\times \mathbb{S}^1$ and the Lie algebra of vector fields on $\mathbb{S}^1$
is the Virasoro algebra \cite{ward}. But the vector field in the
Virasoro algebra is not a Hamiltonian vector field because
$\mathbb{S}^1$ is not a symplectic manifold. So it is required that
the Gibbons-Hawking metric resides in a four-dimensional symplectic
manifold.}

\subsection{LeBrun metric}

As was pointed out in section 3, the LeBrun metric (\ref{lebrun}) is
a solution of the Einstein-Maxwell equation. Therefore it is
nontrivial to solve (\ref{e-preserve}) to determine $\lambda$. Hence
we will show some details of our calculation. For the given frame
(\ref{frame}), it is also necessary to take a frame rotation like
(\ref{so4-rot}) but with a modified form
\begin{equation}\label{rebrun-rot}
    \left(
      \begin{array}{c}
        E'_3 \\
      E'_4 \\
      \end{array}
    \right) = \left(
                \begin{array}{cc}
                  \cos \frac{\phi}{2} & - \sin \frac{\phi}{2} \\
                  \sin \frac{\phi}{2} & \cos \frac{\phi}{2} \\
                \end{array}
              \right) \left(
      \begin{array}{c}
        E_3 \\
      E_4 \\
      \end{array}
    \right),
\end{equation}
where the angle variable $\phi$ along the U(1) fiber is defined by
\begin{equation}\label{u1-angle}
    \phi \equiv \tau + \int^z A_3 (\mathbf{x}) dz.
\end{equation}
We still keep (1-2)-plane unchanged. For $a=1,2$, it is possible to
solve (\ref{e-preserve}) with $\lambda = w^{\frac{1}{2}}
e^{\frac{u}{2}}$. But, for $a=3,4$, there is an extra term with the
final result:
\begin{equation}\label{div-34}
\begin{array}{c}

    \nabla \cdot E'_3 = \frac{w^{-\frac{1}{2}}}{2} \cos \frac{\phi}{2}
    \Big(\partial_z \log  w e^{u} -(w- \partial_z u) \Big), \\
\nabla \cdot E'_4 = \frac{w^{-\frac{1}{2}}}{2} \sin \frac{\phi}{2}
    \Big(\partial_z \log  w e^{u} -(w- \partial_z u) \Big).
\end{array}
\end{equation}
Thus, in order to cancel the extra term, it is required to choose
$\lambda$ properly without affecting the result for $a=1,2$. It
turns out that it can be done by introducing a dipole-like object
given by
\begin{eqnarray}\label{owl-lambda}
    \lambda &=& \exp \Big(- \mathbf{k} \cdot \frac{1}{2}\int_0^1 d\sigma
    \frac{d \mathbf{x}(\sigma)}{d\sigma}
    \big(w- \partial_z u \big) (\mathbf{x}(\sigma)) \Big) w^{\frac{1}{2}}
e^{\frac{u}{2}} \nonumber \\
&\equiv& \Psi(\mathbf{x}) w^{\frac{1}{2}} e^{\frac{u}{2}}
\end{eqnarray}
where the path $P$ is taken along $\mathbb{R}^3$ with the vector
$\mathbf{k} = (0,0,1)$. We will simply call $\Psi(\mathbf{x})$ an
open Wilson line because $w- \partial_z u$ is a gauge field as can
be seen from (\ref{1-form-a}).

One can check that the result (\ref{owl-lambda}) is consistent with
the previous ones. It is obvious that the frame rotation
(\ref{rebrun-rot}) reproduces (\ref{so4-rot}) for the real heaven
case with $A_3 (\mathbf{x}) = 0$ and the relation $w =
\partial_z u$ trivializes the open Wilson line in
(\ref{owl-lambda}). For the Gibbons-Hawking metric with $u=0$ but
$A_3 (\mathbf{x}) \neq 0$, we don't have to take a frame rotation at
the outset. Nevertheless we can solve (\ref{e-preserve}) in a
rotated frame like (\ref{rebrun-rot}) too. In such a rotated frame,
we get an extra factor $w$ in (\ref{div-34}) due to the frame
rotation which must be canceled out by the open Wilson line in
$\lambda$. In this respect, the LeBrun metric (\ref{lebrun}) is a
kind of mixture of these two metrics.

The vector fields $V_a = \lambda E_a \in \Gamma(TM)$ for the LeBrun
metric (\ref{lebrun}) are then given by (after dropping the prime)
\begin{eqnarray} \label{lebrun-frame}
\begin{array}{l}
V_1 = \Psi(\mathbf{x}) \Big(\frac{\partial}{ \partial x} - A_1
\frac{\partial}{ \partial \tau} \Big), \\
    V_2 = \Psi(\mathbf{x}) \Big(\frac{\partial}{ \partial y} -
    A_2 \frac{\partial}{ \partial \tau} \Big), \\
V_3 = e^{\frac{u}{2}}  \Psi(\mathbf{x}) \left( \cos \frac{\phi}{2}
\Big(\frac{\partial}{ \partial z} - A_3 \frac{\partial}{ \partial \tau} \Big)
- w \sin \frac{\phi}{2} \frac{\partial}{ \partial \tau} \right), \\
V_4 = e^{\frac{u}{2}}  \Psi(\mathbf{x}) \left( \sin \frac{\phi}{2}
\Big(\frac{\partial}{ \partial z} - A_3 \frac{\partial}{ \partial \tau} \Big)
+ w \cos \frac{\phi}{2} \frac{\partial}{ \partial \tau} \right).
\end{array}
\end{eqnarray}
Note that $\partial_x \Psi(\mathbf{x}) = \partial_y \Psi(\mathbf{x})
= 0$ and $\partial_z \Psi(\mathbf{x}) = - \frac{1}{2} (w- \partial_z
u) \Psi(\mathbf{x})$. A straightforward calculation shows that the
vector fields in (\ref{lebrun-frame}) are not divergence-free unlike
the previous self-dual metrics. Instead they obey the relation
\begin{equation}\label{divergence-lebrun}
\begin{array}{l}
\partial_\mu V_1^\mu = \partial_\mu V_2^\mu = 0, \\
\partial_\mu V_3^\mu = (w- \partial_z u) \Psi(\mathbf{x}) e^{\frac{u}{2}} \cos
\frac{\phi}{2}, \\
\partial_\mu V_4^\mu = (w- \partial_z u) \Psi(\mathbf{x}) e^{\frac{u}{2}} \sin
\frac{\phi}{2}.
\end{array}
\end{equation}
One may notice that the divergence-free condition is violated even
for the Gibbons-Hawking metric after the frame rotation
(\ref{rebrun-rot}) although the real heaven case was not affected by
it. However, it should not be taken as a surprise because this
divergence-free condition is not preserved under a general internal
rotation of basis vectors.

It will be worthwhile to recall that the bottom-up approach
implicitly assumes the on-shell condition.  This means that we have
to assume the Toda equation (\ref{toda}) and its linearization
(\ref{l-toda}) for the solution (\ref{lebrun}) from the outset. As
we remarked in the second paragraph of section 3, a linear deviation
from a Toda point, $u(\mathbf{x}) = u_t(\mathbf{x}) +
w(\mathbf{x})$, still satisfies the Toda equation (\ref{toda}) as
long as $u_t(\mathbf{x})$ is a solution of (\ref{toda}). Using this
property, we may choose a particular path $P$ in order to define the
open Wilson line (\ref{owl-lambda}) such that, at the end point of
the path where $\mathbf{x} (\sigma = 1) = (x,y,z)$,
\begin{equation}\label{on-shell}
    \partial_x \Big( u(\mathbf{x}) - \int^z w(\mathbf{x}) dz \Big)
    = \partial_y \Big( u(\mathbf{x}) - \int^z w(\mathbf{x}) dz \Big) = 0,
    \qquad \partial_z u(\mathbf{x}) = w(\mathbf{x}).
\end{equation}
Such a path $P$ can be chosen with impunity because the path $P$
obeying (\ref{on-shell}) is consistent with the Toda equation
(\ref{toda}) and its linearization (\ref{l-toda}):
\begin{equation}\label{leb-path}
    \partial_x^2 u + \partial_y^2 u + \partial_z^2 e^u =
    \int^z \Big( \partial_x^2 w + \partial_y^2 w + \partial_z^2 (e^u
    w) \Big)dz = 0.
\end{equation}
For this reason, we call a path $P$ obeying (\ref{on-shell})
on-shell path.\footnote{We have chosen the end point $\mathbf{x}
(\sigma = 1) = (x,y,z)$ for the differentiation of
$\Psi(\mathbf{x})$. We may equally choose the other end point
$\mathbf{x} (\sigma = 0)$ for a differentiation point, as well.
Hence the on-shell condition (\ref{on-shell}) actually must be
imposed on both ends. Then it means that the U(1) fiber represented
by $w - \partial_z u$ is pinched off at two end points of the open
Wilson line in (\ref{owl-lambda}), which is very similar to the
situation of Figure 1 in \cite{bnw11}. Therefore the open Wilson
line represents a two-cycle in the LeBrun metric.} Adopting the same
prescription of path ordering in noncommutative gauge theory
\cite{owl-1,owl-2,owl-3}, we will consider the Lie algebra of vector
fields consisting of all local functions attached at one end of the
open Wilson line with the on-shell condition (\ref{on-shell}) being
satisfied. Then (\ref{divergence-lebrun}) suggests that, on the
on-shell path, the vector fields $V_a$ for the LeBrun metric are
actually divergence-free.

The LeBrun metric (\ref{lebrun}) is a four-dimensional K\"ahler
metric with a vanishing Ricci scalar. As we explained in the last
part of section 2, the emergent gravity implies that such a
scalar-flat K\"ahler manifold can emerge from pure Maxwell theory on
a symplectic manifold. If the LeBrun metric is an example of such a
case, it has to satisfy (\ref{scalar-dual}). Now we will show that
the LeBrun metric (\ref{lebrun}) certainly obeys the scalar-flat
condition (\ref{scalar-dual}).

For this purpose, let us determine the coefficients ${g_{ab}}^c$ in
the infinite-dimensional Lie algebra (\ref{vstr-eq}). In this
calculation, we will use the result (\ref{1-form-a}) for the U(1)
field strength but we will not assume the on-shell condition
(\ref{on-shell}) which will be imposed at the very last stage. A
straightforward though tedious calculation shows that
\begin{eqnarray} \label{lie-1}
[V_1, V_2] &=& \Psi(\mathbf{x}) e^{\frac{u}{2}} \partial_z \log (w
e^u) \Big(\sin \frac{\phi}{2} V_3 - \cos \frac{\phi}{2} V_4 \Big), \\
\label{lie-2}
[V_3, V_4] &=& - \Psi(\mathbf{x}) e^{\frac{u}{2}} \partial_z \log (w
e^u) \Big(\sin \frac{\phi}{2} V_3 - \cos \frac{\phi}{2} V_4 \Big), \\
\label{lie-3}
[V_1, V_3] &=& \frac{1}{2} \Psi(\mathbf{x}) e^{\frac{u}{2}} (w -
\partial_z u) \cos \frac{\phi}{2} V_1 +
\frac{1}{2} \Psi(\mathbf{x}) \Big( \partial_x u V_3 + \int^z \partial_y w dz
 V_4 \Big)\nonumber \\
&& +  \Psi(\mathbf{x}) \Big( \partial_x \log w \sin \frac{\phi}{2} -
\partial_y \log w \cos \frac{\phi}{2} \Big)
\Big(\sin \frac{\phi}{2} V_3 - \cos \frac{\phi}{2} V_4 \Big), \\
\label{lie-4}
[V_2, V_4] &=& \frac{1}{2} \Psi(\mathbf{x}) e^{\frac{u}{2}} (w -
\partial_z u) \sin \frac{\phi}{2} V_2 +
\frac{1}{2} \Psi(\mathbf{x}) \Big( \partial_y u V_4 + \int^z \partial_x w dz
 V_3 \Big)\nonumber \\
&& +  \Psi(\mathbf{x}) \Big( \partial_x \log w \sin \frac{\phi}{2} -
\partial_y \log w \cos \frac{\phi}{2} \Big)
\Big(\sin \frac{\phi}{2} V_3 - \cos \frac{\phi}{2} V_4 \Big), \\
\label{lie-5}
[V_1, V_4] &=& \frac{1}{2} \Psi(\mathbf{x}) e^{\frac{u}{2}} (w -
\partial_z u) \sin \frac{\phi}{2} V_1 +
\frac{1}{2} \Psi(\mathbf{x}) \Big( \partial_x u V_4 - \int^z \partial_y w dz
 V_3 \Big)\nonumber \\
&& -  \Psi(\mathbf{x}) \Big( \partial_x \log w \cos \frac{\phi}{2} +
\partial_y \log w \sin \frac{\phi}{2} \Big)
\Big(\sin \frac{\phi}{2} V_3 - \cos \frac{\phi}{2} V_4 \Big), \\
\label{lie-6}
[V_2, V_3] &=& \frac{1}{2} \Psi(\mathbf{x}) e^{\frac{u}{2}} (w -
\partial_z u) \cos \frac{\phi}{2} V_2 +
\frac{1}{2} \Psi(\mathbf{x}) \Big( \partial_y u V_3 - \int^z \partial_x w dz
 V_4 \Big)\nonumber \\
&& +  \Psi(\mathbf{x}) \Big( \partial_x \log w \cos \frac{\phi}{2} +
\partial_y \log w \sin \frac{\phi}{2} \Big)
\Big(\sin \frac{\phi}{2} V_3 - \cos \frac{\phi}{2} V_4 \Big).
\end{eqnarray}
From the above results, one can easily read off the coefficients
${g_{ab}}^c$ in the Lie algebra (\ref{vstr-eq}). Using the
definition (\ref{rho-psi}), one can deduce the following relations
\begin{eqnarray} \label{s-flat-lebrun}
\begin{array}{l}
\rho_1 + \Psi_1 = \partial_x \Big( u(\mathbf{x}) - \int^z
w(\mathbf{x}) dz \Big) \Psi(\mathbf{x}), \\
\rho_2 + \Psi_2 = \partial_y \Big( u(\mathbf{x}) - \int^z
w(\mathbf{x}) dz \Big) \Psi(\mathbf{x}), \\
\rho_3 + \Psi_3 = -(w- \partial_z u) \Psi(\mathbf{x}) e^{\frac{u}{2}} \cos
\frac{\phi}{2}, \\
\rho_4 + \Psi_4 = -(w- \partial_z u) \Psi(\mathbf{x})
e^{\frac{u}{2}} \sin \frac{\phi}{2}.
\end{array}
\end{eqnarray}
Interestingly, the divergence equation (\ref{divergence-lebrun})
indicates that $\rho_3 + \Psi_3 = - \partial_\mu V_3^\mu$ and
$\rho_4 + \Psi_4 = - \partial_\mu V_4^\mu$. In (\ref{s-flat-lebrun})
and above equations, we are implicitly assuming the prescription of
path ordering described below (\ref{leb-path}) to attach local
functions at one end of $\Psi(\mathbf{x})$.

As we have justified before, we can choose a path $P$ in order to
satisfy the on-shell condition (\ref{on-shell}) to define the open
Wilson line (\ref{owl-lambda}). Strictly speaking, it is actually
required because the two functions $u(\mathbf{x})$ and
$w(\mathbf{x})$ must satisfy (\ref{toda}) and (\ref{l-toda}),
respectively. So far we have not imposed the on-shell condition
(\ref{on-shell}) anywhere. After applying the on-shell condition
(\ref{on-shell}) to (\ref{s-flat-lebrun}), we can immediately deduce
that the scalar flat condition
\begin{equation}\label{s-flat-asd}
    \rho_a = - \Psi_a
\end{equation}
is truly satisfied. This fact demonstrates that the LeBrun metric
(\ref{lebrun}) can arise from pure Maxwell theory on a
four-dimensional symplectic manifold whose equations of motion are
given by (\ref{map-eom}).

We will not try to solve (\ref{lebrun-frame}) to obtain symplectic
gauge fields for the LeBrun metric. It may be premature before
getting them for the Gibbons-Hawking metric (\ref{gh-metric})
because the LeBrun metric (\ref{lebrun}) contains (\ref{gh-metric})
and (\ref{real-heaven}) as particular cases. But we will get back to
the problem in the near future.

\section{Conclusion}

Let us recapitulate the lesson perceived from the bottom-up approach
of emergent gravity. In the top-down approach of emergent gravity,
we have symplectic gauge fields (or noncommutative gauge fields at
very short distances) and their dynamical equations of motion. The
most accessible frame in this case is the Darboux frame where
symplectic gauge fields are defined by solving (\ref{darboux-tr}).
But we are not compelled to reside in the Darboux frame as we
already emphasized in the footnote \ref{diff-darboux}. In principle
we can formulate the gauge theory in an $\Omega$-frame (using the
notation of the footnote \ref{diff-darboux}). The gauge theory in
this case will be described by the Poisson bracket
(\ref{t-poisson-bracket}) with a nontrivial Poisson tensor
$\Theta^{\mu\nu}(x)$ like (\ref{g-poisson}) and the corresponding
symplectic gauge fields are then defined by (\ref{symp-gen}). It was
previously argued (see the paragraph around (2.26) in
\cite{hsy-jhep09}) that the resulting gauge theory is equivalent to
the gauge theory on a curved space with a canonical Poisson tensor.
Under both circumstances (either with a nontrivial Poisson tensor on
a flat space or with a canonical Poisson tensor on a curved space),
the construction of the full noncommutative gauge theory is a
challenging problem. Thus the Darboux frame provides the most
rudimentary gadget to formulate emergent gravity. But a caveat is
that we cannot make a direct comparison with a gravitational metric
since the gravitational metric is represented in a general
coordinate system which is not necessarily in the Darboux frame, as
we already emphasized in section 1. If we could formulate gauge
theory and its emergent gravity in a general $\Omega$-frame, it
would be possible to directly get gravitational metrics in Einstein
gravity from the top-down approach.

In the bottom-up approach of emergent gravity explored in this
paper, we start with a gravitational metric given on a Riemannian
manifold $M$. We can solve (\ref{gauge-phi}) to determine the Weyl
factor $\lambda$ for a given metric and then identify the vector
fields in (\ref{ham-vc}) via the relationship (\ref{v-e}). We found
that the Weyl factor $\lambda$ for a general metric contains a
dipole-like object (which we dubbed an open Wilson line according to
the similarity appearing in noncommutative gauge theory).
Nevertheless, either gravitational metrics or tetrads are still
described by local functions according to the relation (\ref{v-e}).
An intriguing point is that it seems unnecessary to introduce the
dipole-like Weyl factor for gravitational instantons and we believe
that this may be applicable to all kinds of gravitational
instantons. However it turned out that symplectic gauge fields in a
general coordinate system are nonlocal functions even in commutative
limit but with a Poisson structure defined thereto. The appearance
of nonlocality may be expected due to the following reasons. In
noncommutative gauge theories, local gauge invariant observables do
not exist since we can effect a spatial translation by a gauge
transformation \cite{owl-3}. The interrelation between a gauge
transformation and a spatial translation still persists in the
commutative limit as we remarked in the footnote \ref{gauge-trans}.
Therefore we cannot construct a local gauge invariant observable
using symplectic gauge fields. This is also consistent with the idea
of emergent gravity. In general relativity there exist no local
gauge invariant observables either, as translations are equivalent
to general coordinate transformations. Thereby, from this point of
view, the emergence of dipole-like objects may be quite natural when
we try to define symplectic gauge fields from a gravitational
metric.

In spite of some difficulty to treat nonlocal objects such as
(\ref{symp-gauge}) and (\ref{owl-lambda}), the bottom-up approach of
emergent gravity nicely confirms the results of the top-down
approach and elucidates many important aspects on emergent gravity
as was summarized above. For example, the bottom-up approach renders
a novel verification of the equivalence between gravitational
instantons and symplectic U(1) instantons \cite{hsy-epl09}. In
particular, the real heaven case presents a paragon of the bottom-up
approach by successfully producing the solution (\ref{rh-gsol}) of
the BPS equation (\ref{rh-bps}). If one tries to solve the equation
(\ref{rh-bps}) directly, it would be difficult to embody a solution.
We think it already demonstrates a sound aspect of the bottom-up
approach for emergent gravity. In addition to the explicit solution,
a more noteworthy success is to verify that the LeBrun metric
(\ref{lebrun}) is a solution of pure Maxwell theory on a
four-dimensional symplectic manifold whose equations of motion are
given by (\ref{map-eom}). It also constitutes a nontrivial check of
the formula (\ref{emergent-einstein}) derived in \cite{hsy-jhep09}.
Therefore, if we can extract symplectic gauge fields from the vector
fields in (\ref{lebrun-frame}) by solving (\ref{vec-comp}), it will
constitute a very general class of solutions for noncommutative
gauge theory and quantum gravity. We hope to open that direction
with the work \cite{our-future}.

One may be tempted to apply the bottom-up approach of emergent
gravity to (Euclidean) Schwarzschild black hole. The Euclidean
Schwarzschild black hole solution describes a Ricci-flat manifold
\cite{hawking-ebh}. But it is not a K\"ahler manifold. So it does not
admit a natural symplectic structure available in the $\Omega$-frame
(\ref{gsym-str}). The best alternative choice is to utilize the
(anti-)self-dual harmonic two-forms on the space (see eq.(3) in
\cite{bh-harmonic}) and define a Poisson algebra determined by the
self-dual harmonic two-form. However a magnetic mass (and an
electric mass) at the origin seems to bring about the violation of
Jacobi identity of the underlying Poisson algebra similar to the
situation of a charged quantum particle in the presence of a
magnetic monopole \cite{jackiw}. Therefore the Schwarzschild black
hole remains a challenging goal for the top-down as well as the
bottom-up approaches of emergent gravity.

So far we have implicitly assumed that fluctuations $F$ in
(\ref{symp-omega}) have no homogeneous sink on vacuum. In other
words, we have exclusively considered local fluctuations so that
$|F| \to 0$ at asymptotic infinity. In this case the Darboux frame
is defined by a coordinate transformation $\phi \in
\mathrm{Diff}(M)$ obeying $\phi^* (B + F) = B$ as in
(\ref{darboux-tr}). But we may consider a general kind of
fluctuations allowing a homogeneous condensate on vacuum. This means
that fluctuations $F$ will change even the asymptotic vacuum
structure and so the Darboux transformation $\phi
\in \mathrm{Diff}(M)$ instead is defined by $\phi^* (B + F) = B +
\langle F(|x| \to \infty) \rangle_{\mathrm{vac}} \equiv B'$.
If $\mathrm{rank}(B) = \mathrm{rank}(B')$, we may introduce a
nowhere vanishing function $f$ such that $B' = fB$. Therefore, to
describe such situation, we need to introduce an almost symplectic
manifold $(M, \Omega)$ where the two-form $\Omega$ is nondegenerate
but not necessarily closed and, in particular, is locally conformal
to a symplectic form $\omega$. Such an almost symplectic manifold is
known as a locally conformal symplectic manifold
\cite{lcs-lee,conf-vaisman}. On a locally conformal symplectic manifold $(M,
\Omega)$, there exists an open covering ${U_\alpha}$ of $M =
\bigcup_\alpha U_\alpha$ and a smooth positive function $f_\alpha$
on each $U_\alpha$ such that $f_\alpha\Omega|_{U_\alpha} \equiv
\Omega_\alpha$ is symplectic on $U_\alpha$. This is equivalent to the
existence of a closed one-form $\eta$, the so-called Lee form
\cite{lcs-lee}, such that
\begin{equation}\label{lee-form}
    d \Omega = \eta \wedge \Omega.
\end{equation}
When $\eta$ vanishes identically, we recover the symplectic two-form
$\Omega$. And any Hamiltonian vector field $X$ on a locally
conformal symplectic manifold satisfies \cite{conf-vaisman}
\begin{equation}\label{lcs-vector}
    \mathcal{L}_X \Omega_\alpha = k \Omega_\alpha
\end{equation}
with a constant $k$. It turns out \cite{hsy-inflation} that it is
necessary to introduce such a locally conformal symplectic structure
to describe the epoch of cosmic inflation of our universe. If so,
the locally conformal symplectic structure might play an important
role for the birth and the evolution of our Universe.

In conclusion, we are yet to invite the most important two
players--the Schwarzschild black hole and the cosmic inflation--to
the league of emergent gravity. Confrontations with them will
certainly help us lift the veil of quantum gravity.

\acknowledgments This work was supported by the National Research Foundation
of Korean (NRF) grant funded by the Korea government (MEST) through
the Center for Quantum Spacetime (CQUeST) of Sogang University with
grant number 2005-0049409. The research of HSY was also supported by
Basic Science Research Program through the National Research
Foundation of Korea (NRF) funded by the Ministry of Education,
Science and Technology (2011-0010597). This work was performed
during SGL's visit to CQUeST. He thanks CQUeST for hospitality and
support during that period.

\end{document}